 \documentclass[12pt]{article}
 \usepackage{graphicx}
 \usepackage{color}
 \usepackage{amssymb, latexsym, amsmath}
    \usepackage{kformat}

 \newcommand{\be}{\begin{equation}}
 \newcommand{\ee}{\end{equation}}
 \newcommand{\ba}{\begin{eqnarray}}
 \newcommand{\ea}{\end{eqnarray}}
 \newcommand{\inc}{{\it i}}
 \newcommand{\bra}{\langle}
 \newcommand{\ket}{\rangle}
 \newcommand{\efbold}{\mbox{{\boldmath $\vec f$}}}
 
 \newcommand{\erbold}{\mbox{{\boldmath $\vec r$}}}
 
 \newcommand{\mubold}{\mbox{{\boldmath $\vec \mu$}}}

 \newcommand{\wbold}{\mbox{{\boldmath $\vec w$}}}
  \newcommand{\dotmubold}{
  \stackrel{\textbf{.}}{
  {\bf{\mbox{\boldmath
  ${\boldmath\vec{\boldmath{\mu}}}$}}}}}
 \newcommand{\ddoterbold}{
 {\ddot{\mbox{${\vec{\bf{r}}}$}}}
 }

 \newcommand{\cc}{\textrm{c}}
 \newcommand{\s}{\textrm{s}}
 \newcommand{\yr}{\textrm{yr}}

\begin{document}
 \title{
 {\Large{\textbf{Long-term evolution of orbits about a
 precessing\footnote{~We use the term ``precession" in its general meaning,
 which includes any change of the instantaneous spin axis. So generally
 defined precession embraces the entire spectrum of spin-axis variations --
 from the polar wander and nutations through the Chandler wobble through the
 equinoctial precession.} ~oblate planet:\\
 3. A semianalytical and a purely numerical approach.}
            }}}
 \author{
  {\Large{Pini Gurfil}}\\
 {\small{Faculty of Aerospace Engineering, Technion, Haifa 32000 Israel}}\\
 {\small{e-mail: pgurfil @ technion.ac.il~}},\\
 ~\\
 {\Large{Val\'ery Lainey}}\\
 {\small{Observatoire Royal de Belgique, Avenue Circulaire 3, Bruxelles 1180
 Belgium,}}\\
 {\small{IMCCE/Observatoire de Paris, UMR 8028 du CNRS, 77 Avenue
 Denfert-Rochereau, Paris 75014 France}}\\
 {\small{e-mail: Valery.Lainey @ imcce.fr~}},\\
~\\
 and\\
 ~\\
 {\Large{Michael Efroimsky}}\\
 {\small{US Naval Observatory, Washington DC 20392 USA}}\\
 {\small{e-mail: me @ usno.navy.mil~}}
 }
 \maketitle
 \begin{abstract}
 Construction of an accurate theory of orbits about a precessing and
 nutating oblate planet, in terms of osculating elements defined in a frame
 associated with the equator of date, was started in Efroimsky \& Goldreich
 (2004) and Efroimsky (2004, 2005, 2006a,b). Here we continue this line of research
 by combining that analytical machinery with numerical tools. Our model
 includes three factors: the $\,J_2\,$ of the planet, its nonuniform
 equinoctial precession described by the Colombo formalism, and the
 gravitational pull of the Sun. This semianalytical and
 seminumerical theory, based on the Lagrange planetary equations for
 the Keplerian elements, is then applied to Deimos on very long time scales
 (up to 1 billion of years). In parallel with the said
 semianalytical theory for the Keplerian elements defined in the
 co-precessing equatorial frame, we have also carried out a completely
 independent, purely numerical, integration in a quasi-inertial Cartesian
 frame. The results agree to within fractions of a percent,
 thus demonstrating the applicability of our semianalytical model over
 long timescales.

 Another goal of this work was to make an independent check of whether the
 equinoctial-precession variations predicted for a rigid Mars by the Colombo
 model could have been sufficient to repel its moons away from the equator.
 An answer to this question, in combination with our knowledge of the current
 position of Phobos and Deimos, will help us to understand whether the Martian
 obliquity could have undergone the large changes ensuing from the said model
 (Ward 1973; Touma \& Wisdom 1993, 1994; Laskar \& Robutel 1993), or whether
 the changes ought to have been less intensive (Bills 2006, Paige et al. 2007).
 It has turned out that, for low initial inclinations, the orbit inclination
 reckoned from the precessing equator of date is subject only to small
 variations. This is an extension, to non-uniform equinoctial precession
 given by the Colombo model, of an old result obtained by Goldreich (1965)
 for the case of uniform precession and a low initial inclination. However,
 near-polar initial inclinations may exhibit considerable variations for
 up to $\pm 10$ deg in magnitude. This result is accentuated when the
 obliquity is large. Nevertheless, the analysis confirms that an oblate
 planet can, indeed, afford large variations of the equinoctial
 precession over hundreds of millions of years, without repelling its
 near-equatorial satellites away from the equator of date: the satellite
 inclination oscillates but does not show a secular increase. Nor does it
 show secular decrease, a fact that is relevant to the discussion of the
 possibility of high-inclination capture of Phobos and Deimos.

 \end{abstract}

 \section{Introduction}

 \subsection{Statement of purpose}

 The goal of this paper is to explore, by two very different methods,
 inclination variations of a solar-gravity-perturbed satellite orbiting
 an oblate planet subject to nonuniform equinoctial precession. This
 nonuniformity of precession is
 caused by the presence of the other planets. Their gravitational pull
 entails precession of the circumsolar orbit of our planet; this entails
 variations of the solar torque acting on it; these torque variations make
 the planet's equinoctial precession nonuniform; and this nonuniformity, in
 its turn, influences the behaviour of the planet's satellites. This
 influence is feeble, and we trace with a high accuracy whether it results,
 over aeons, in purely periodic changes in inclination or can accumulate to
 secular changes.

 This work is but a small part of a larger project whose eventual goal is to
 build up a comprehensive tool for computation of long-term orbital evolution
 of satellites. Building this tool, block by block, we are beginning with only
 three components -- the planet's oblateness, the direct pull of the Sun on
 the satellite and the planet's precession. These phenomena bare a marked
 effect on the evolution of the orbit. In our subsequent publications, we
 shall incorporate more effects into our model -- 
 the triaxiality, and the bodily tides.

 \subsection{Motivation}

 One motivation for this work stems from our intention to carry out an
 independent check of whether the equinoctial-precession changes predicted
 for a rigid Mars by the Colombo model could have been sufficient to repel
 its moons away from the equator. An answer to this question, in combination
 with our knowledge of the current position of Phobos and Deimos, will help
 us to understand whether the Martian obliquity variations could indeed have
 undergone the large variations resulting from the Colombo model, or whether
 the actual variations ought to have smaller magnitude. Such a check is
 desirable because the current, Colombo-model-based theory of equinoctial
 precession (Ward 1973; Touma \& Wisdom 1993, 1994; Laskar \& Robutel 1993),
 incorporates several approximations. First, the Colombo equation is derived
 under the assertion that the planet is rigid and is always in its
 principal spin state, the angular-momentum vector staying parallel to the
 angular-velocity one. Second, this description, being only a model, ignores
 the possibility of planetary catastrophes that might have altered the
 planet's spin mode. Third, this description ignores that sometimes even weak
 dissipation (caused, for example, by tides) may be sufficient to quell chaos
 and regularise the motion, which may be the case of Mars (Bills 2006). Fourth,
 it still remains a matter of controversy as to whether the observed pattern of
 small craters on Mars confirms (Hartmann 2007) or disproves (Paige et al. 2007)
 the strong climatic variations predicted in Ward (1974, 1979). All this
 motivates us to come up with a test based on the necessity to reconcile the
 variations of spin with the present near-equatorial positions of Phobos and
 Deimos. The fact that
 both moons found themselves on near-equatorial orbits, in all likelihood,
 billions of years ago,\footnote{~Phobos and Deimos give every appearance of
 being captured asteroids of the
 carbonaceous chondritic type, with cratered surfaces older than $\,\sim
 10^9\,$ years (Veverka 1977; Pang et al. 1978; Pollack et al. 1979; Tolson
 et al. 1978). If they were captured by gas drag (Burns 1972, 1978), this
 must have occurred early in the history of the solar system while the gas
 disk was substantial enough. Kilgore, Burns, and Pollack (1978) have
 demonstrated numerically that a gas envelope extending to about ten
 Martian radii, with a density of $\,5\,\times\,10^{-5}\,$g/cm$^3$ at the
 Martian surface, could have been capable of capturing satellites of radii
 about $\,10$km.
 At that stage of planetary formation, the spin of the forming planet would
 be perpendicular to the planet's orbit about the Sun -- i.e., the obliquity
 would be small and the gas disk would be nearly coplanar with the planetary
 orbit. Energetically, a capture would likely be equatorial. This is most
 easily seen in the context of the restricted three-body problem. The
 surfaces of zero velocity constrain any reasonable capture to occur from
 directions near the inner and outer collinear Lagrange points (Szebehely
 1967, Murison 1988), which lie in the equatorial plane. Also, a somewhat
 inclined capture would quickly be equatorialised by the gas disk. If the
 capture inclination is too high, the orbital energy is then too high to
 allow a long enough temporary capture, and the object would hence not
 encounter enough drag over a long enough time to effect a permanent capture
 (Murison 1988).
 Thus, Phobos and Deimos were likely (in as much as we can even use that
 term) to have been captured into near-equatorial orbits.} and that
 both are currently located within less than $\,2\,$ degrees from the
 equator, is surely more than a mere coincidence. An elegant but
 sketchy calculation by Goldreich (1965) demonstrated that the orbits
 of initially near-equatorial satellites remain close to the equator of date
 for as long as some simplifying assumptions remain valid. As explained in
 Efroimsky (2004, 2005), these assumptions are valid over time scales not
 exceeding $\,100\,$ million years, while at longer times a more careful
 analysis is required. Its goal will be to explore the
 limits for the possible secular drift of the satellite orbits away from the
 evolving equator of date. Through comparison of these limits with the
 present location of the Martian satellites, we shall be able to impose
 restrictions upon the long-term spin variations of Mars. If, however, it
 turns out that near-equatorial satellites can, in the face of large
 equinoctial-precession variations, remain for billions of years
 close to the moving equator of date, then we shall admit that Mars' equator
 could indeed have precessed through billions of years in the manner predicted
 by the Colombo approximation.

 The second motivation for our study comes from the ongoing discussion of whether
 the Martian satellites might have been captured at high inclinations, their
 orbits having gradually approached the equator afterwards. While a comprehensive
 check of this hypothesis will need a more detailed model -- one that will include
 Mars' triaxiality, the tidal forces, (Lainey, Gurfil, \& Efroimsky 2008), and
 perhaps other perturbations -- the first, rough sketch of this test can be carried
 out with only $\,J_2\,$, the Sun, and the equinoctial precession taken into account.
 We perform such a rough check for a hypothetical satellite that has all the
 parameters of Deimos, except that its initial inclination is $\;89^o\;$.

 \subsection{Mathematical tools}

 The first steps toward the analytical theory of orbits about a precessing and
 nutating Earth were undertaken almost half a century ago by Brouwer (1959),
 Proskurin \& Batrakov (1960), and Kozai (1960). The problem was considered,
 in application to the Martian satellites, by Goldreich (1965) and, in regard to
 a circumlunar orbiter, by Brumberg, Evdokimova, \& Kochina (1971). The latter
 two publications addressed the dynamics as seen in a non-inertial frame
 co-precessing with the planet's equator of date. The
 analysis was carried out in terms of the so-called ``contact" Kepler elements,
 i.e., in terms of the Kepler elements satisfying a condition
 different from that of osculation. Modeling of perturbed trajectories by
 sequences of instantaneous ellipses (or hyperbolae) parameterised with such
 elements is sometimes very convenient mathematically (Efroimsky 2006c). However, the
 physical interpretation of such solutions is problematic, because
 instantaneous conics defined by nonosculating elements are nontangent to
 the trajectory. Though over restricted time scales the secular parts of the
 contact elements may well approximate the secular parts of their osculating
 counterparts (Efroimsky 2004, 2005), the cleavage between them may grow at longer
 time scales. This is the reason why a practically
 applicable treatment of the problem must be performed in the language of
 osculating variables.

 \subsection{The plan}

The analytical theory of orbits about a precessing oblate primary,
in terms of the Kepler elements defined in a co-precessing (i.e.,
related to the equator of date) frame, was formulated in Efroimsky
(2006a,b) where the planetary equations were approximated by
neglecting some high-order terms and averaging the others. This way,
from the exact equations for osculating elements, approximate
equations for their secular parts were obtained. We shall borrow
those averaged Lagrange-type planetary equations, shall incorporate
into them the pull of the Sun, and shall numerically explore their
solutions. This will give us a method that will be semianalytical
and seminumerical. We shall then apply it to a particular setting --
evolution of a Martian satellite and its reaction to the long-term
variations of the spin state of Mars. Our goal will be to explore
whether the spin-axis variations predicted for a rigid Mars permit
its satellites to remain close to the equator of date for hundreds
of millions through billions of years. In case the answer to this
question turns out to be negative, it will compel us to seek
nonrigidity-caused restrictions upon the spin variations. Otherwise,
the calculations of the rigid-Mars inclination variations will
remain in force (and so will the subsequent calculations of Mars
obliquity variations); this way, the theory of Ward (1973, 1974,
1979, 1982), Touma \& Wisdom (1993, 1994), and Laskar \& Robutel
(1993) will get a model-independent confirmation.

 \section{Semianalytical treatment of the problem}

To understand the evolution of a satellite orbit about a precessing
planet, it is natural to model it with elements defined in a
coordinate system associated with the equator of date, i.e., in a
frame co-precessing (but not co-rotating) with the planet. A
transition from an inertial frame to the co-precessing one is a
perturbation that depends not only upon the instantaneous position
but also upon the instantaneous velocity of the satellite. It has
been demonstrated by Efroimsky \& Goldreich (2004) that such
perturbations enter the planetary equations in a nontrivial way: not
only they alter the disturbing function (which is the negative
Hamiltonian perturbation) but also they endow the equations with
several extra terms that are not parts of the disturbing function.
Some of these nontrivial terms are linear in the planet's precession
rate $\;\mubold\;$, some are quadratic in it; the rest are linear in
its time derivative $\;\dotmubold\;$. The inertial-forces-caused
addition to the disturbing function (i.e., to the negative
Hamiltonian perturbation) consists of a term linear and a term
quadratic in $\;\mubold\;$. (See formulae (53 - 54) in Efroimsky
(2004, 2005) or formulae (1) and (6) in Efroimsky (2006a,b).) The
essence of approximation elaborated in {\emph{Ibid.}} was to neglect
the quadratic terms and to substitute the terms linear in
$\;\mubold\;$ and $\;\dotmubold\;$ with their secular parts
calculated with precision up to $\;e^3\;$, inclusively.

 \subsection{Equations for the secular parts of osculating elements
             defined in a co-precessing reference frame.}

We shall begin with five Lagrange-type planetary equations for the
secular parts of the orbital elements defined in a frame
co-precessing with the equator of date. These equations, derived in
Efroimsky (2004, 2005), have the following form:
 \ba
 \frac{da}{dt}\,=\;-\;2\;\frac{\dot{\mu}_{\perp}}{n}\;\,a\;\,
 \left(1\,-\,e^2\right)^{1/2}\;\;\;,\;\;~~~~~~~~~~~~~~~~~~~~~
 ~~~~~~~~~~~~~~~~~~~~~~~~~
 \label{1}
  \ea
  \begin{eqnarray}
 \frac{de}{dt}\,=\;\frac{5}{2}\;\frac{\dot{\mu}_{\perp}}{n}\;e
 \;\left(\,1\;-\;e^2\,\right)^{1/2}~~~,~~~~~~~~~~~~~~~~~~~~~~~
 ~~~~~~~~~~~~~~~~~~~~~~~~~
 \label{2}
  \end{eqnarray}
  \begin{eqnarray}
 \frac{d\omega}{dt}=\frac{3}{2}\frac{n\,J_2}{\left(1-e^2\right)^2}\left(
 \frac{\rho_e}{a}\right)^2\left(\frac{5}{2}\,\cos^2\inc-\frac{1}{2}\right)-
 {\mu_{\perp}}+{\mu}_n\,\cot\inc\,
 - \frac{\cos\inc}{n a^2(1-e^2)^{1/2} \sin \inc }\;\langle\dotmubold\left(
 -\efbold\times\frac{\partial \efbold}{\partial \inc} \right)\rangle
 \;,~~
 \label{3}
  \end{eqnarray}
 ~\\
 \ba
 \nonumber
 \frac{d
 \inc}{dt}\;=~-\,\mu_1\,\cos\Omega\,-\,\mu_2\,\sin\Omega\,~~~~~~~~~
 ~~~~~~~~~~~~~~~~~~~~~~~~~~~~~~~~~~~~~~~~~~~~~~~~~~~~~~~~~~~~~~~~~~
 ~~~~~~~
 \ea
 \ba
 ~~~~\,+\,\frac{\cos \inc}{n a^2(1 - e^2)^{1/2} \sin
 \inc}\;
 \langle \;\dotmubold\left(\,-\,\efbold\times
 \frac{\partial \efbold}{\partial \omega}  \right)\,
 \rangle\;-\,\frac{1}{na^2\,(1-e^2)^{1/2}\,\sin \inc
 }\;\langle \, \dotmubold\, \left(\,-\,\efbold\,\times\,
 \frac{\partial \efbold}{\partial \Omega}  \right)\;
 \rangle\;\;\;{,}\;\;\;\;~~~
 \label{4}
  \ea
 ~\\
 ~\\
 \ba
 \frac{d\Omega}{dt}\,=\,-\,\frac{3}{2}\,n\,J_2\,\left(\,\frac{\rho_e}{a}\,\right)^2\;
 \frac{\cos\inc}{\left(1\;-\;e^2\right)^2}\;-\;\frac{\mu_n}{\sin \inc}\;
 +\;\frac{1}{n\; a^2\;(1\;-\;e^2)^{1/2}\;\sin \inc }\;\langle\,\dotmubold\left(
 \,-\,\efbold\times\frac{\partial\efbold}{\partial\inc}\right)\,\rangle\,\;\;,\;\;\;\;~~~
 \label{5}
  \ea
  ~\\
 The number of equations is five, because one element, $\,M_o\,$, was
 excluded by averaging of the Hamiltonian perturbation and of the inertial
 terms emerging in the right-hand sides. In the equations,
 $\,n\,\equiv\,\sqrt{G\left(m_{primary}\,+\,m_{secondary}\right)/a^3~}\,$,
 while $\,\efbold (\,t\,;\;a\,,\;e\,,\;\inc\,,\;\omega\,,\;\Omega\,,\;M_o)\,
 $ is the implicit function that expresses the unperturbed two-body dependence
 of the position upon the time and Keplerian elements. Vector $\,\mubold
 \,$ denotes the total precession rate of the planetary equator (including
 all spin variations -- from the polar wander and nutations through the
 Chandler wobble through the equinoctial precession through the
 longest-scale spin variations caused by the other planets' pull), while $\,
 \mu_1\,,\,\mu_2\,,\,\mu_3\,$ stand for the components of $\,\mubold\,$ in a
 co-precessing coordinate system $\,x\,,\,y\,,\,z\,$, the axes $\,x\,$ and $
 \,y\,$ belonging to the equator-of-date plane, and the longitude of the
 node, $\,\Omega\,$, being reckoned from $\,x\,$:
 \ba
 \mubold\;=\;\mu_1\,{\bf{\hat x}}\;+\;\mu_2\;
 {\bf{\hat y}}\;+\;\mu_3\;{\bf{\hat
 z}}\;\;\;,\;\;\;\;{\mbox{where}}\;\;\;\;\mu_1\;=\;{\dot{I}_p}\;\;\;,\;\;\;\;
 \mu_2\;=\;{\dot{h}_p}\,\sin I_p\;\;\;,\;\;\;\;\mu_3\;=\;{\dot{h}_p}\,\cos
 I_p\;\;\;,\;\;\;
 \label{6}
 \ea
 $I_p\,,\;h_p\;$ being the inclination and the longitude of the node of the equator of date
 relative to that of epoch, and a dot standing for a time
 derivative. The quantity $\;\mu_{\perp}\;$ is a component of
 $\;\mubold\;$ directed along the instantaneous orbital momentum
 of the satellite, i.e., perpendicular to the instantaneous
 osculating Keplerian ellipse. This component is expressed with
 \ba
 \mu_{\perp}\;\equiv\;\mubold\,\cdot\,\wbold\;=\;\mu_1\;\sin \inc\;\sin \Omega\;-
 \;\mu_2\;\sin \inc\;\cos \Omega\;+\;\mu_3\;\cos \inc\;\;\;\;,
 \label{7}
 \ea
the unit vector
 \ba
 {\wbold}\;=\;{\bf\hat{x}}\;\sin \inc\;\sin \Omega\;-
 \;{\bf\hat{y}}\;\sin
 \inc\;\cos \Omega\;+\;{\bf\hat{z}}\;\cos
 \inc\;\;~~~~~~~~~~~~~~~~~~~
 \label{8}
 \ea
standing for the unit normal to the instantaneous osculating
ellipse. Be mindful that $\;\dot{\mu}_{\perp}\;$ is defined not as
$\;d(\mubold\cdot\wbold)/dt\;$ but as
 \ba
 \dot{\mu}_{\perp}\;\equiv\;\dotmubold\cdot\,\wbold\;=\;
 \dot{\mu_1}\;\sin \inc\;\sin \Omega\;-
 \;\dot{\mu_2}\;\sin \inc\;\cos \Omega\;+\;\dot{\mu_3}\;\cos \inc
 \;\;\;.
 \label{9}
 \ea
 The quantity $\;\mu_n\;$ is a component pointing within the satellite's orbital
 plane, in a direction orthogonal to the line of nodes of the satellite orbit
 relative to the equator of date:
 \ba
 \nonumber
 \mu_n\;=\;\;-\;{\mu}_1\;\sin \Omega\;\cos\inc\,+\,
 {\mu}_2\;\cos \Omega\;\cos \inc\,+\,{\mu}_3\;\sin
 \inc~~~~~~~~~~~~~~~~~~ \\
 \label{10}\\
 \nonumber
 =\;-\;\dot{I}_p\;\sin \Omega\;\cos\inc\,+\,
 \dot{h}_p\,\sin I_p\;\cos \Omega\;\cos \inc\,+\,
 \dot{h}_p\,\cos I_p\;\sin\inc\;\;\;.
 \ea
 Its time derivative taken in the frame of reference co-precessing
 with the equator of date is:
  \ba
  \nonumber
 {\dot{\mu}}_n\;=\;-\;\dot{\mu}_1\;\sin \Omega\;\cos\inc\,+\,\dot{\mu}_2\;
 \cos\Omega\;\cos \inc\,+\,\dot{\mu}_3\;\sin \inc~~~~~~~~~~~~~~~~~~~~~~~~~~~~~~~~
 ~~~~~~~~~~~~~~~~~~~~~~~~~~~~ \\
 \label{}\\
 \nonumber
 =\;-\;\ddot{I}_p\;\sin \Omega\;\cos\inc\,+\,
 \left(\,
 {\ddot{h}}_p\,\sin I_p\,+\,{\dot{h}}_p\,{\dot{I}}_p\,\cos I_p
 \,\right)
 \;\cos \Omega\;\cos \inc\,+\,
 \left(\,
 {\ddot{h}}_p\,\cos I_p\,-\,{\dot{h}}_p\,{\dot{I}}_p\,\sin I_p
 \,\right)
 \;\sin
 \inc\;\;.\;\;
 \label{11}
 \ea
As shown in Efroimsky (2006a,b), the $\;\dotmubold$-dependent terms, emerging
in equations (\ref{1} - \ref{5}), are expressed with
 \ba
 \nonumber
 \bra\;\dotmubold\cdot\left(\;-\;\efbold\;\times\;
 \frac{\partial {\efbold}}{\partial \inc}\;
 \right)\;\ket\;=~~~~~~~~~~~~~~~~~~~~~~~~~~~~~~~~~~~~~~~~~~~~~~~~~~~~~~~~~~~~~~~~~~~~~~~~~~~\\
 \nonumber\\
 \nonumber\\
 \nonumber
 \frac{a^2}{4}\;\left\{\;
 \dot{\mu}_1\;\left[\;-\;\left(2\,+\,3\,e^2\right)
 \,\cos\Omega\,+\,5\,e^2\;\left(\cos\Omega\;\cos 2\omega\,-\,\sin
 \Omega\;\sin 2\omega\;\cos\inc \right)
 \;\right]\;+ \right.\\
 \nonumber\\
 \nonumber\\
 \nonumber
 \dot{\mu}_2\;\left[\;-\;\left(2\,+\,3\,e^2\right)
 \,\sin\Omega\,+\,5\,e^2\;\left(\sin\Omega\;\cos 2\omega\,+\,\cos
 \Omega\;\sin 2\omega\;\cos\inc \right)
 \;\right]\;+\,\\
 \nonumber\\
 \nonumber\\
 \left. \dot{\mu}_3\;\left[\;5\;e^2\;\sin
 2\omega\;\sin\inc
 \;\right]\;\right\}~~~~~~~~~~~~~~~~~~~~~~~~~~~~~~~~~~~~~~~~~~~~~~~~~~~~~~~~~~~~~~\,
 \label{12}
 \ea
  \ba
 \bra\;\dotmubold\,\cdot\,
 \left(\,-\,\efbold\;\times\;\frac{\partial \efbold}{\partial
 \omega}\;\right)\;\ket\,=\,
 -\;\frac{a^2}{2}\;
 \left(2\;+\;3\;e^2\right)\;\left(\;\dot{\mu}_1\;\sin\inc\;\sin\Omega
 \;-\;\dot{\mu}_2\;\sin\inc\;\cos\Omega
 \;+\;\dot{\mu}_3\;\cos\inc
 \;\right)~~~,~~~~
 \label{13}
 \ea
 \ba
 \nonumber
 \bra\;\dotmubold\cdot\left(\;-\;\efbold\;\times\;
 \frac{\partial {\efbold}}{\partial \Omega}\;
 \right)\;\ket\;=
 ~~~~~~~~~~~~~~~~~~~~~~~~~~~~~~~~~~~~~~~~~~~~~~~~~~~~~~~~~~~~~~~~~~~~~~~~~~~~~~~~~~~~~~~~~~~~\\
 \nonumber\\
 \nonumber\\
 \nonumber
 \frac{a^2}{4}\;\left\{\;
 \dot{\mu}_1\;\sin\inc\;
\left[\;-\;\left(2\,+\,3\,e^2\right)\;\sin\Omega\;\cos\inc\;+\;5\;e^2\;\left(\,
\cos \Omega\;\sin 2\omega\,+\,\sin\Omega\;\cos 2\omega\;\cos
\inc \,\right) \;\right] \right. \\
 \nonumber\\
 \nonumber\\
 \nonumber
 +\;\dot{\mu}_2\;\sin\inc\;\left[\;
 \left(2\,+\,3\,e^2\right)\;\cos\Omega\;\cos\inc\;+\;5\,e^2\;\left(
 \sin\Omega\;\sin 2\omega\;-\;\cos\Omega\;\cos 2\omega\;\cos\inc
 \right)
 \;\right]~~~~\\
 \nonumber\\
 \nonumber\\
 \left. \;-\;\dot{\mu}_3\;\left[\;
 \left(2\,+\,3\,e^2\right)\,\left(2\;-\;\sin^2\inc\;\right)\;+\;
 {5}\,e^2\,\sin^2\inc\;\cos 2\omega
 \;\right]
 \;\right\}
 ~~~~~~~~~~~~~~~~~~~~~~~~~~~~~~~~~~~
 \label{14}
 \ea
 ~\\
 To integrate equations (\ref{1} - \ref{5}), with expressions
 (\ref{12} - \ref{14}) inserted therein, we shall need to know,
 at each step of integration, the components of $\,\mubold\,$
 and $\,\dotmubold\,$.

 \subsection{Calculation of the components of $\;\mubold\;$ and $\;\dotmubold\;$.}\label{subsec:Calcofcomp}

 At each step of our integration, the components of $\;\mubold\;$ and
 $\;\dotmubold\;$ will be calculated in the Colombo approximation.
 Physically, the essence of this approximation is
two-fold: first, the solar torque acting on the planet is replaced by its
average over the year; and, second, the precessing planet is assumed to be
always in its principal spin state. While a detailed development (based on the
work by Colombo (1966)) may be found in the Appendix to Efroimsky (2006a,b),
here we shall provide a concise list of resulting formulae to be used.

The components of $\;\mubold\;$ are connected, through the medium
of (\ref{6}), with the inclination and the longitude of the node
of the moving planetary equator, $\;I_p\;$ and $\;h_p\;$, relative
to some equator of epoch. These quantities and their time
derivatives are connected with the unit vector $\;{\bf\hat k}\;$
aimed in the direction of the major-inertia axis of the planet:
 \ba
  {\bf{\hat{k}}}\;=\;\left(\;\sin I_p\,\sin h_p\;\;,\;\;\;-\,\sin I_p\,\cos h_p
  \;\;,\;\,\;\cos I_p\;\right)^{^T}\;\;\;.
 \label{18}
 \ea
This unit vector and its time derivative
 \ba
  \frac{d\bf{\hat{k}}}{dt}=\left(
  {\dot{I}}_p\,\cos I_p\,\sin h_p\,+\,{\dot{h}}_p\,\sin I_p\,\cos h_p
  \;\,,\;\;\,
  -\,{\dot{I}}_p\,\cos I_p\,\cos h_p\,+\,{\dot{h}}_p\,\sin I_p\,\sin h_p
  \;\,,\;\,\;
  -\,{\dot{I}}_p\,\sin I_p\right)^{^T}\;\;\;,\;
 \label{19}
 \ea
depend, through the Colombo equation
 \ba
 \frac{d{\bf\hat k}}{dt}\;=\;\alpha\;\left({\bf{\hat{n}}}\cdot{\bf{\hat{k}}}
 \right)\,\left({\bf{\hat{k}}}\times{\bf{\hat{n}}}  \right)\;\;\;,
 \label{20}
 \ea
upon the unit normal to the planetary orbit,
 \ba
 {\bf\hat n}\;=\;\left(\;\sin I_{orb}\,\sin
\Omega_{orb}\;\;,\;\;\;-\,\sin I_{orb}\,\cos \Omega_{orb}
  \;\;,\;\,\;\cos I_{orb}\;\right)^{^T}\;\;\;,
 \label{21}
 \ea
 $\Omega_{orb}\;$ and $\;I_{orb}\;$ being the node and inclination of the orbit
 relative to some fiducial fixed plane, and $\,\alpha\,$ being a parameter
 proportional to the oblateness factor $\,J_2\,$. In our computations, we
 employed the present-day\footnote{~
 An accurate treatment shows that
 due to the precession of the Martian orbit $\,\alpha\,$ exhibits quasi-periodic
 variations of about $\,3\,\%\,$ over long time scales. Neglecting this detail
 in the current work, we shall take it into account at the further stage of our
 project (Lainey, Gurfil \& Efroimsky 2008).} value of $\,\alpha\,$ -- see Table 5 below.
 We chose this value because it was the one used by Ward (1973, 1974), and we wanted
 to make sure that our plot for obliquity evolution, Fig.~\ref{fig2_paper}, coincided
 with that of Ward (1974).

 To find the components of $\,\mubold\,$,
 one must know the time evolution of $\,h_p\,$ and $\,I_p\,$, which can be determined
by solving a system of three differential equations (\ref{20}), with
$\,\Omega_{orb}\,$ and $\,I_{orb}\,$ being some known functions of
time. These functions may be computed via the auxiliary variables
 \ba
 q\;=\;\sin I_{orb}\;\sin \Omega_{orb}\;\;\;,\;\;\;\;\;
 p\;=\;\sin I_{orb}\;\cos \Omega_{orb}\;\;\;,
 \label{22}
 \ea
whose evolution will be given by the formulae:
 \ba
 q\;=\;\sum_{j=1}^{\infty} \,N_j\,\sin \left(s_j't\,+\,\delta_j
 \right)\;\;\;,
 \label{23}
 \ea
 \ba
 p\;=\;\sum_{j=1}^{\infty} \,N_j\,\cos \left(s_j't\,+\,\delta_j
\right)\;\;\;.
 \label{24}
 \ea
 The following choice of the values of the amplitudes, frequencies,
 and phases will make equations (\ref{22} - \ref{24}) render
 $\;\Omega_{orb}\;$ and $\;I_{orb}\;$ relative to the solar system's
 invariable plane (Ward, 1974):~\footnote{~~Ward (1974) has calculated
 the values of the coefficients $N,\,s',\delta$ based on the work of
 Brouwer \& van Woerkom (1950) who had calculated the values of
 $\,p\,$ and $\,q\,$ relative to the ecliptic plane of 1950. Ward
 (1974) transformed Brouwer \& van Woerkom's values to the solar
 system's invariable plane. Be mindful that, no matter what the
 reference plane, both Ward (1974) and Brouwer \& van Woerkom (1950)
 used the same epoch, J1950.}

\begin{table}[h]
\begin{center}
\begin{tabular}{ c c c c }
  \hline
  $j$ & $N_j$ & $s_j'\; [{\rm arcsec/yr}]$ & $\delta_j'\; [{\rm deg}]$ \\
  \hline\hline
  1 & 0.0018011 & -5.201537 & 272.06 \\[0.5ex]
  2 & 0.0018012 & -6.570802  & 210.06 \\[0.5ex]
  3 & -0.0358910 & -18.743586 & 147.39 \\[0.5ex]
  4 & 0.0502516 & -17.633305 & 188.92 \\[0.5ex]
  5 & 0.0096481 & -25.733549 & 19.58 \\[0.5ex]
  6 & -0.0012561 & -2.902663 &  207.48 \\[0.5ex]
  7 & -0.0012286 & -0.677522 & 95.01 \\
  \hline
  \end{tabular}
  \caption{Numerical values used in Ward's model of the inclination
and node of the Martian orbit}
\label{table1}
\end{center}
\end{table}

 The development by Ward (1974) is
 limited in precision. A more accurate treatment was offered by Laskar (1988). At the future stages of our
 project, when developing a detailed physical model of the satellite motion, we shall employ
 Laskar's results. In the current paper, we are checking if the orbital
 averaging of the precession-caused terms is permissible at large time scales. Hence, for the
 purpose of this check, as well for the qualitative estimate of the long-term behaviour
 of the satellites, we need a realistic, not necessarily highly accurate, scenario of the precession
 variations.

%
%
%
%
%

 \subsection{The Goldreich approximation}

 The above semianalytical treatment not only yields plots of the time
 dependence of the mean elements but also serves as a launching pad for
 analytical approximations. For example, an assumption of $\,a\,$ and $\,
 e\,$ being constant, and a neglect of the $\,\dotmubold$-dependent terms
 in (\ref{4} - \ref{5}), as well as of the term $\;\;\mu_n/\sin\inc\;\;$
 in (\ref{5}), gives birth to the Goldreich (1965) approximation:
 \ba
 \frac{da}{dt}\; &=& \;0~~~,
 \label{25}\\
 \frac{de}{dt}\; &=& \;0~~~,
 \label{26}\\
 \frac{d i}{dt}\; &=& \;-\;\mu_1 \;\cos\Omega\;-\;\mu_2\;\sin \Omega~~~,
 \label{27}\\
 \frac{d\Omega}{dt}\;&=&\;-\;\frac{3}{2}\;n\;J_2\;\left(\frac{\rho_e}{a}\right)^2\;\frac{\cos
 \inc }{(1\,-\,e^2)^2}~~~,
 \label{28}\\
 \frac{d\omega}{dt}~&=&~\frac{3}{4}~n~J_2~\left(\frac{\rho_e}{a}\right)^2~\frac{5~\cos^2 \inc~
 -~1}{(1\,-\,e^2)^2}\;+\;\frac{\mu_n\;\cos \inc}{\sin \inc
}\;-\;\mu_{\perp}~~~,
 \label{29}
 \ea
  the equinoctial precession being assumed uniform:
  \begin{eqnarray}
    \dot{I}_p \; & = & \; 0
    \label{30}\\
    \dot{h}_p \; & = & \; -\alpha\cos I_p
    \label{31}\\
    \mu_1 \; & = &\; 0
    \label{32}\\
    \mu_2 \; & = & \; \dot{h}_p\sin I_p
    \label{33}\\
    \mu_3 \; & = & \; \dot{h}_p \;\cos I_p\;\;\;.
    \label{34}
  \end{eqnarray}
 For the details of Goldreich's approximation see also subsection 3.3 in
 Efroimsky (2005).

 \subsection{The Gravitational Pull of the Sun}

Reaction of a satellite on the planetary-equator precession is, in a
way, an indirect reaction of the satellite on the presence of the
Sun and the other planets. Indeed, the pull of the other planets
makes the orbit of our planet precess, which entails variations in
the Sun-produced gravitational torque acting on the planet. These
variations of the torque, in their turn, lead to the variable
equinoctial precession of the equator, precession ``felt" by the
satellite. It would be unphysical to consider this, indirect effect
of the Sun and the planets upon the satellite, without taking into
account their direct gravitational pull. In this subsection, we
shall take into account the pull of the Sun, which greatly dominates
that of the planets other than the primary.

In what follows, $\,m\,$ and $\,m'\,$ will be the masses of the
satellite and the Sun, correspondingly, $\,\erbold\,$ and
$\,\erbold'\,$ will stand for the planetocentric positions of the
satellite and the Sun, $\,S\,$ will signify the angle between these
vectors. Then the Sun-caused perturbing potential $\,R_S\,$, acting
on the satellite, will assume the form of
 \begin{equation}
 R_S=Gm^\prime\left(\frac{1}{|\erbold^{\ \prime}-\erbold|}-\frac{\erbold^{\ \prime}\cdot \erbold}{r^{\prime
 3}}\right)=\frac{Gm^\prime}{r^\prime}\left(\frac{r^\prime}{|\erbold^{\ \prime}-\erbold|}-\frac{r\cos S}{r^\prime}\right)
 \label{RS}
 \end{equation}
This can be expanded in a usual manner over the Legendre polynomials
of the first kind. Since $\,r'\,\gg\,r\,$, we shall take only the
first term in the series:\footnote{~~This is justified since the
next term in the Legendre series is about 2 orders of magnitude
smaller than the potential variations generated by the precession.}

 \begin{eqnarray}
 \label{}
    R_S &\approx&\frac{Gm^\prime}{r^\prime}\left[1+\frac{r}{r^\prime}P_1(\cos S)+\left(\frac{r}{r^\prime}\right)^2P_2(\cos
    S)-\frac{r\cos S}{r^\prime}\right]
 \end{eqnarray}

 As the term ${Gm^\prime}/{r^\prime}$ is not dependent of the satellite's elements, one has only to consider the restricted
 potential

 \begin{eqnarray}
 \label{}
    R_S^1 =
    \frac{Gm'}{r\,'}\left[\left(\frac{r}{r\,'}\right)^2 P_2(\cos
    S)\right]\;
     \approx \frac{n\,'^{\,2}\;a^{\,2}}{2}\left(\frac{a\,'}{r\,'}\right)^3(3\;\cos^2 S\,-\,1)\;\;\;,
 \end{eqnarray}
$n\,'\,$ and $\,a\,'\,$ being the mean motion and the semi-major
axis of the Sun.

To obtain the Lagrange-type equations for the third-body
perturbation, we must first derive an expression for the angle
$\,S$. To that end, define the directional cosines
$\,\xi\,\equiv\,\hat{\vec P}\cdot{\hat{{\vec{r}}\,}}'\,$ and
$\,\theta\,\equiv\,\hat{\vec Q}\cdot {\hat{\vec r}\,}'\,$, where
$\,{\hat{\vec r}\,}'\,$ is a unit vector pointing from the planet
toward the Sun, while $\,\hat{\vec P}\,$ and $\,\hat{\vec Q}\,$ are
unit vectors of a perifocal coordinate system associated with the
osculating orbital plane of the satellite, with $\hat{\vec P}$
pointing to the instantaneous periapse and $\hat{\vec Q}$ being
orthogonal to $\hat{\vec P}$. Assuming that the planet's orbit about
the Sun is circular, we arrive at
 \begin{equation}
 \label{r1}
    \xi\; =\; \cos\omega\;\cos(\Omega\,-\,M')\;-\;\cos \inc\;\sin
    \omega\;\sin(\Omega\,-\,M')\;\;\;,\;\;\;\;
 \end{equation}
 \begin{equation}
 \label{r2}
    \theta\; = \;-\;\sin\omega\;\cos(\Omega\,-\,M')\;-\;\cos \inc
    \;\cos\omega\;\sin(\Omega\,-\,M')\;\;\;,
 \end{equation}
where $\,M'\,$ is the mean anomaly of the Sun in the planetocentric
frame. With aid of these relations, the angle $\,S\,$ may be written
down as
 \begin{equation}
 \label{r3}
 \cos S\; =\; \xi\;\cos \nu\;+\;\theta\; \sin \nu\;\;\;,
 \end{equation}
$\nu\,$ being the true anomaly of the satellite in the
planetocentric coordinate system. Substituting (\ref{r1}) and
(\ref{r2}) into (\ref{r3}), and averaging over the satellite's mean
anomaly, we arrive at the Lagrange-type equations:
 \begin{subequations}
 \label{3body}
 \begin{eqnarray}
  \frac{da}{d\tau} &=& 0
  \label{vavila}\\
  \nonumber\\
  \frac{de}{d\tau} &=& 10\,e\,\sqrt{1-e^2}\,\left[\sin^2\inc\;\sin
 2\omega+(2-\,\sin^2 \inc)\;\sin 2\;\omega\;\cos 2\tilde{\Omega}+2\;\cos
 \inc\;\cos 2\omega\;\sin 2\tilde\Omega\right]~~~~~
 \label{gloria}\\
  \nonumber\\
  \frac{d\inc}{d\tau} &=&  -\;\frac{2\;\sin \inc}{\sqrt{1-e^2}}\left\{5\;e^2\;\cos \inc\;\sin 2\omega\;\;(1\,-\,\cos 2\tilde\Omega)\,-\left[2\,+\,e^2(3\,+\,5\;\cos 2\omega)\right]\;\sin 2\tilde\Omega\right\} \\
  \nonumber\\
  \nonumber
  \frac{d\omega}{d\tau} &=&  \frac{2}{\sqrt{1-e^2}}\,\left\{4+e^2-\,5\;\sin^2\inc\,+\,5\,(\sin^2\inc\,-\,e^2
  )\;\cos2\omega\,+\,5\,(e^2-2)\;\cos \inc\;\sin 2\omega\;\sin 2\tilde\Omega \right.\\
  \nonumber \\
                        &~&~~~~~~~+\;\left.\left[5\,(2\,-\,e^2\,-\;\sin^2 \inc)\;\cos 2\omega\,-\,2\,-\,3\,e^2\,+\,5\;\sin^2 \inc\right]\;\cos2\tilde\Omega\right\}\\
  \nonumber\\
  \frac{d\tilde\Omega}{d\tau} &=&  -\,\kappa\,-\,\frac{2}{\sqrt{1-e^2}}\,\left\{\left[2+e^2(3-5\;\cos 2\omega)\right]\,(1-\;\cos 2\tilde\Omega)\,\cos \inc-5e^2\;\sin 2\omega\;\sin 2\tilde\Omega\right\}
 \end{eqnarray}
 \end{subequations}
where we used the following set of notations:
 \begin{equation}
 \label{}
 \tau\; \equiv \;\beta \;n\;(t\,-\,t_0)\;\;\;,\;\;\;\;\;\;\;
 \beta = \frac{3\,m'\,a^3}{16\,M\,a'^{\textstyle{^{\,3}}}}
 \end{equation}
 \begin{equation}
 \label{}
 \kappa\;\equiv\; \frac{16\,n}{3\,n'}\left(1\,+\,
 \frac{M}{m'}\right)
 \end{equation}
 \ba
 \tilde \Omega \equiv \Omega - \lambda'\;\;\;,
 \label{}
 \ea
 $\lambda'\,$ being the mean longitude of the Sun in the planet's frame, and $M$ being the mass of Mars.

An additional averaging can be performed over the motion of the
planet about the Sun. Mathematically, this is the same as averaging
over the motion of the Sun about the planet -- in both cases
averaging over $\,\lambda'\,$ is implied. This averaging (Innanen et
al. 1997) will simplify (\ref{3body}) into
\begin{subequations}
\label{3body1}
\begin{eqnarray}
  \frac{da}{dt} &=& 0 \\
  \frac{de}{dt} &=& 10\;e\;\sqrt{1-e^2}\;\beta\;n\;\sin^2\inc\;\sin 2\omega \\
  \frac{di}{dt} &=&  -\;\frac{10\;e^2\;\sin \inc}{\sqrt{1-e^2}}\;\beta\;n\;\cos \inc\;\sin 2\omega \\
  \frac{d\omega}{dt} &=&  \frac{2}{\sqrt{1-e^2}}\;\beta\;n\;\left[\,4\;+\;e^2\;-\;5\;\sin^2\inc\;+\;5\;(\sin^2
  \inc\,-\;e^2)\;\cos 2\omega\right]\\
  \frac{d\Omega}{dt} &=&   -\;\frac{2}{\sqrt{1-e^2}}\;\beta\;n\;\left[\,2\,+\,e^2\,(3\,-\,5\;\cos 2\omega)\,\right]\;\cos \inc
\end{eqnarray}
\end{subequations}
It is important to note that the calculations leading to
Eqs.~(\ref{3body}), and to their double-averaged version,
Eqs.~(\ref{3body1}), were performed in the Martian-orbital, and not
Martian-equatorial frame, without taking either the Martian
obliquity or precession into consideration. Had we taken into
account the precession, we would get, on the right-hand side of
(\ref{3body1}) resonances between the motion of the Sun relative to
the planet and the equinoctial precession. Since the time scale of
the former exceeds, by orders of magnitude, the time scale of the
latter, we may safely omit such resonances. This justifies our
neglect of the frame precession in the above calculation.

\begin{center}
\begin{figure}[h]
  \includegraphics[width=7.0in]{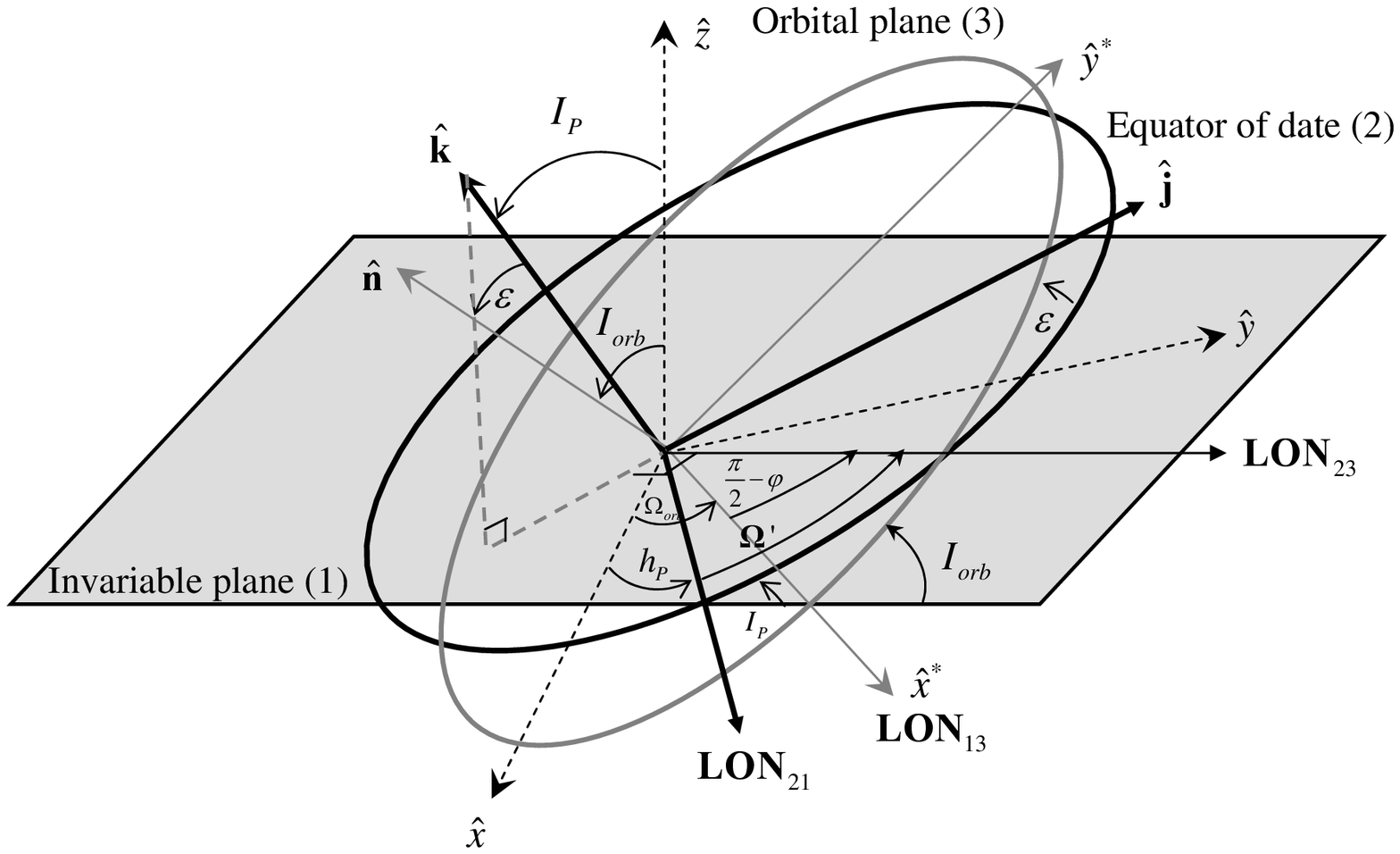}\\
  \caption{The geometry of the precessing Mars. The Martian spin axis, $\bf
\hat{k}$, is perpendicular to the equator of date, and the normal to
the orbital plane (ecliptic of date), $\bf \hat{n}$. The Martian
obliquity, $\epsilon$, is the angle between these two vectors.
  }
  \label{planessmallbw}
\end{figure}
\end{center}

However, the omission of the obliquity may have a serious effect on
the results. Thus, we shall generalize Eqs.~(\ref{3body}) so as to
include the effect of the solar inclination and node in a
Martian-centric frame. This generalized model based on the
celebrated works by Kozai (1959) and Cook (1962) gives us:
\begin{subequations}
\label{3body2}
\begin{eqnarray}
 \frac{da}{dt} &=& 0 \\
  \frac{de}{dt} &=& -\frac{15 n'^2 e\sqrt{1-e^2}}{4 n}\left [2AB\cos(2\omega)-(A^2-B^2)\sin(2\omega)\right]\\
  \frac{di}{dt} &=&  \frac{3 n'^2 C}{4 n\sqrt{1-e^2}}\left\{A\left [2+3e^2+5e^2\cos(2\omega)\right]+5Be^2\sin(2\omega)\right\}\label{42c0}\\
  \frac{d\omega}{dt} &=&   -\dot\Omega\cos i+\frac{3 n'^2\sqrt{1-e^2}}{2  n}\left[5AB\sin(2\omega)+\frac{5}{2}(A^2-B^2)\cos(2\omega)-1+\frac{3(A^2+B^2)}{2}\right]\nonumber \\
  &+& \frac{15 n'^2 a(A\cos \omega+B\sin\omega)}{4 nea'}\left[1-\frac{5}{4}(A^2+B^2)\right]\\
  \frac{d\Omega}{dt} &=&   \frac{3n'^2C}{4 n\sqrt{1-e^2}\sin
  i}\left\{5Ae^2\sin(2\omega)+ {B}\left[2+3e^2-5e^2\cos(2\omega)\right]\right\}
\end{eqnarray}
\end{subequations}
where
\begin{subequations}
\begin{eqnarray}
 A &=&  \cos(\Omega-\Omega')\cos(\lambda')+\cos(i')\sin(\lambda')\sin(\Omega-\Omega') \\
 \nonumber\\
 \nonumber\\
  B &=& \cos i[-\sin(\Omega-\Omega')\cos(\lambda')+\cos(i')\sin(\lambda')\cos(\Omega-\Omega')]\nonumber \\
  \nonumber\\
  &+&\sin i \sin(i')\sin(\lambda')  \\
  \nonumber\\
  \nonumber\\
 C &=& \sin i[\cos \lambda'\sin(\Omega-\Omega')-\cos(i')\sin(\lambda')\cos(\Omega-\Omega')]\nonumber \\
 \nonumber\\
 &+&\cos i\sin i'\sin \lambda'
 \nonumber\\
\end{eqnarray}
\end{subequations}
Here, $i'$ and $\Omega'$ are the inclination and right ascension of
the ascending node of the Solar orbit in the Martian equatorial
frame, respectively. The doubly averaged equations, obtained after
averaging over the Sun's mean anomaly for a single period, are given
by
\begin{subequations}
\label{3body3}
\begin{eqnarray}
 \nonumber\\
 \frac{da}{dt} &=& 0 \\
 \nonumber\\
 \nonumber\\
  \frac{de}{dt} &=& -\frac{15 n'^2 e\sqrt{1-e^2}}{4 n}\left [2\overline{AB}\cos(2\omega)-(\overline{A^2}-\overline{B^2})\sin(2\omega)\right]\\
  \nonumber\\
  \nonumber\\
  \frac{di}{dt} &=&  \frac{3 n'^2 }{4 n\sqrt{1-e^2}}\left\{\overline{CA}\left [2+3e^2+5e^2\cos(2\omega)\right]+5\overline{CB}e^2\sin(2\omega)\right\}\label{42c0}\\
  \nonumber\\
  \nonumber\\
  \frac{d\omega}{dt} &=&   -\dot\Omega\cos i+\frac{3 n'^2\sqrt{1-e^2}}{2  n}\left[5\overline{AB}\sin(2\omega)+\frac{5}{2}(\overline{A^2}-\overline{B^2})\cos(2\omega)-1+\frac{3(\overline{A^2}+\overline{B^2})}{2}\right] ~~~~~~~\\
  \nonumber\\
  \nonumber\\
  \frac{d\Omega}{dt} &=&   \frac{3n'^2}{4 n\sqrt{1-e^2}\sin
  i}\left\{5\overline{CA}e^2\sin(2\omega)+ \overline{CB}\left[2+3e^2-5e^2\cos(2\omega)\right]\right\}
  \nonumber\\
  \end{eqnarray}
 \end{subequations}
 the averaged quantities
 $\overline{A^2},\,\overline{B^2},\,\overline{AB},\,\overline{AC},\,\overline{BC}$
 being given by

 \begin{subequations}
 \label{3bodyabc}
 \begin{eqnarray}
 \overline{A^2} &=& \left[  \s_{i'}^2 \cc_{\Omega'}^{2}-0.5\,
  \s_{i'}^2  \right]
 \cc_{\Omega}^{2}+ \s_{i'}^2\s_{\Omega'}\s_{\Omega'} \s_{\Omega} \cc_{\Omega'} \cc_{\Omega} +
0.5-0.5\s_{i'}^2   \cc_{\Omega'}^{2}\\[1.0ex]
\overline{B^2} &=&  \left\{  \left(  \s_{i'}^2  \cc_{\Omega'}^{2}+0.5\s_{i'}^2 \right)  \cc_{\Omega}^{2}- \s_{\Omega'}
 \s_{i'}^{2}   \s_{\Omega} \cc_{\Omega'} \cc_{\Omega} + 0.5\s_{i'}^2 \cc_{\Omega'}^{2}-0.5+ \cc_{i'}^{2}
 \right\}  \cc_{i}^{2}\nonumber\\&+& \left( \cc_{i'} \cc_{\Omega} \cc_{\Omega'} +\cc_{i'} \s_{\Omega} \s_{\Omega'}  \right) \s_{i'} \s_{i} \cc_{i} + 0.5\s_{i'}^2\\[1.0ex]
\overline{AB} & = &  \left\{  \s_{i'}^2 \s_{\Omega'} \cc_{\Omega'} \cc_{\Omega}^{2}+ \left[
 -\s_{i'}^2
\s_{\Omega}  \cc_{\Omega'}^{2}+ 0.5\s_{i'}^2 \s_{\Omega}  \right] \cc_{\Omega}  -0.5\s_{i'}^2 \s_{\Omega'}
\cc_{\Omega'}  \right\} \cc_{i} \nonumber\\&+&
 0.5\cc_{i'}\left( \, \s_{\Omega}
\cc_{\Omega'} -
\cc_{\Omega} \s_{\Omega'}  \right)
\s_{i'} \s_{i}\\[1.0ex]
\overline{AC} &=& 0.5\,\cc_{i'}\left(  \s_{\Omega} \cc_{\Omega'} -
\cc_{\Omega} \s_{\Omega'}  \right)
\s_{i'} \cc_{i} \nonumber\\&+& \left\{  -\s_{i'}^2 \s_{ \Omega'}   \cc_{\Omega'}
\cc_{\Omega}^{2}+ \left[  -\s_{i'}^2 ((\s_\Omega)^2\cc_{\Omega'}^{2}
    ) \right] \cc_\Omega + 0.5\s_{i'}^2 \s_{\Omega'} \cc_{\Omega'}  \right\} \s_{i}\\[1.0ex]
\overline{BC} &=&   \left( \cc_{i'} \cc_{\Omega} \cc_{ \Omega'} +\cc_{i'} \s_\Omega \s_{\Omega'}  \right) \s_{ i'}\cc_{i}^{2}\nonumber\\&+&
 \left\{  \left[  \s_{i'}^2  \left( \cc_{\Omega'}
 \right) ^{2}-0.5\, (s_{i'})^2\right] \cc_{\Omega}^{2}-\s_{i'}^2 \s_{ \Omega'}\s_{\Omega} \cc_{
 \Omega'} \cc_{\Omega} + -0.5\s_{i'}^2  \cc_{ \Omega'}^{2}- \cc_{i'
}^{2}+0.5 \right\} \s_{i} \cc_{i} \nonumber\\&-&   0.5\left(\,\cc_{i'} \s_{\Omega} \s_{\Omega'} +\,\cc_{
 i'} \cc_{\Omega} \cc_{ \Omega'} \right) \s_{i'}
 \end{eqnarray}
 \end{subequations}
where we have used the compact notation
$\cc_{(\cdot)}=\cos(\cdot),\,\s_{(\cdot)}=\sin(\cdot)$. To calculate
the trigonometric functions of $\Omega'$ and $i'$ appearing in
Eqs.~(\ref{3bodyabc}),
 we shall utilise the geometry rendered Figure
 \ref{planessmallbw}. The figure depicts the Martian spin axis, $\bf
\hat{k}$, that is perpendicular to the equator of date, and the
normal to the orbital plane (ecliptic of date), $\bf \hat{n}$. The
Martian obliquity, $\epsilon$, is the angle between these two
vectors, and is calculated based on the Colombo formalism:
 \begin{equation}
 \label{oblo}
    \cos\epsilon = {\bf \hat{k}\cdot \hat{n}} = q\, s_x  + p\,s_y
    +F\, s_z
 \end{equation}
where
\begin{equation}\label{}
    s_x=\sin I_p  \sin h_p\; ,\;\;\;s_y = \sin I_p \cos h_p \;,\;\;\;s_z = \cos
    I_p\;,\;\;\;F = \sqrt{1-p^2-q^2}
\end{equation}
and $\,i'\,=\,\epsilon\,$.

The angle $\Omega'$, lying in the equator of date, is subtended
between the vectors $\textbf{LON}_{12}$ and $\textbf{LON}_{23}$. The
first of these, $\textbf{LON}_{12}$, points along the line-of-nodes
obtained by the intersection of the equator of date and the
invariable plane. This vector must be perpendicular to the plane
defined by $\hat{\vec k}$ and $\bf\hat{z}$ (the normal to the
invariable plane), so that
\begin{equation}\label{}
    \textbf{LON}_{21} \;= \;{\bf\hat{z}}\;\times\;{\hat{\vec k}}\; =\;
    [\;\;s_y\;\;,\;\quad s_x\;\;,\;\quad 0\;\;]^{\textstyle{\;^T}}
\end{equation}
The second vector, $\,\textbf{LON}_{23}\,$, is aimed along the line of
nodes rendered by the intersection of the orbital plane and the
equator of date. Based on the geometry of Fig.~\ref{planessmallbw},
we write:
 \begin{equation}
 \label{}
 \textbf{LON}_{23} \;=\;   \hat{\vec k}\;\times\;\hat {\vec{n} }\;=\;
[\;\;-s_y\,F\,+\,s_z\,p\;\;,\;\quad s_z\,q\,-\,s_x\,F\;\;,\;\quad
-s_x\,p\,+\,s_y\,q\;\;\;]^{\textstyle{\;^T}}
 \end{equation}
By taking the scalar product of $\textbf{LON}_{23}$ and
$\textbf{LON}_{21}$, we derive the direction cosine:
 \ba
 \label{cosOm}
    \cos\Omega'\;=\; \frac{\textbf{LON}_{23}\cdot \textbf{LON}_{21}}{|\textbf{LON}_{23}|\;\;|\textbf{LON}_{21}|}
        \;=\; \frac{\cos I_p\cos i'-\cos I_{orb}}{\sin I_p \sin i'}
\ea
To derive an expression for $\,\sin\Omega'\,$, we shall have to compute
an auxiliary vector, $\,\hat{\vec j}\,$, which lies in the orbital plane
and is normal to both $\,\hat{\vec k}\,$ and $\,\textbf{LON}_{21}\,$ (cf.
Fig.~\ref{planessmallbw}):
 \begin{equation}
 \label{}
 \hat{\vec j} \;\equiv\; \hat{\vec k}\,\times\, \textbf{LON}_{21}\;=\;
 [\;\;-\,s_z\,s_x\;\;,\;\;\quad s_z\,s_y\;\;,\;\;\quad
 s_x^2\,+\,s_y^2\;\;\;]^{\textstyle{\;^T}}
 \end{equation}
 The direction cosine between $\hat{\vec j}$ and $\textbf{LON}_{23}$ is then
 \begin{equation}
 \label{sinOm}
 \cos\left(\;\frac{\pi}{2}\;-\;\Omega'\;\right)\;=\;\frac{\textbf{LON}_{23}\,
 \cdot\,\hat{\vec j}}{|\textbf{LON}_{23}|\;\;\;|\hat{\vec j}|}
 \end{equation}
Upon evaluating Eq.~(\ref{sinOm}) we arrive at
 \begin{equation}\label{}
 \sin\Omega' \;=\; \frac{q\;\cos h_p\;-\;p\;\sin h_p}{\sin i'}
 \end{equation}
Finally, the calculation of the orbital-elements' evolution is
performed via Lagrange-type planetary equations, whose right-hand
sides combine those of (\ref{1}) - (\ref{5}) and (\ref{3body3}).\\

\subsection{The higher-order harmonics, the gravitational pull of the planets,
 the Yarkovsky effect, and the bodily tides}

 In the current paper, we pursue a limited goal of taking into account the
 oblateness of the planet, its nonuniform equinoctial precession, and the
 gravitational pull of the Sun. These three items certainly do not exhaust
 the list of factors influencing the orbit evolution of a satellite.

 Among the factors that we intend to include into the model at the further
 stage of its development are the high-order zonal ($\,J_3\,,\,J_4\,,\,
 J_2^2\,$) and tesseral ($\,C_{22}\,$) harmonics of the planet's gravity
 field, as well as the gravitational pull of the other planets, --
 factors whose role was comprehensively discussed, for example, by Waz (2004).
 We also intend to include the bodily tides (Efroimsky \& Lainey 2007) and
 the Yarkovsky effect (Nezvorn$\acute{\mbox{y}}$ \&
 Vokrouhlick$\acute{\mbox{y}}$ 2007) -- factors whose importance increases
 at long time scales .

  \section{Comparison of a purely numerical and\\ a semianalytical
  treatment of the problem}

 One of our goals is to check the applicability limits (both in terms of the initial
 conditions and the permissible time scales) of our semianalytical model written for
 the osculating elements introduced in a frame co-precessing with the equator of date.
 This check will be performed by an independent, purely numerical, computation that
 will be free from whatever simplifying assumptions (all terms kept, no averaging
 performed.) The straightforward simulation will be carried out in terms of
 Cartesian coordinates and velocities defined in an inertial frame of reference. Both the semianalytical
 calculation of the elements in a co-precessing frame and the straightforward numerical
 integration in inertial Cartesian axes will be carried out for
 Deimos.

 \subsection{Integration by a purely numerical approach.}

 The numerical integration of Deimos' orbit can be performed using Cartesian
 coordinates defined relatively to the Solar system invariable plane. As we also have to compute the Martian
 polar axis motion, there are two vector differential equations to integrate simultaneously.
 One is the Newton gravity law:
 \be
  \ddoterbold
 \,=\;-\;\frac{G\;(M\;+\;m){\erbold}}{r^3}+Gm^\prime\left(\frac{{\erbold}^{\
 \prime}-{\erbold}}{|{\erbold}^{\ \prime}-{\erbold}|^3}-\frac{{\erbold}^{\ \prime}}{r^{\prime 3}}\right)\;+\;G\;(M\;+\;m)\;{\bf{\nabla}}U
 \;\;\;,~~~~~~~~~~~~~~~~~~~~~~
 \label{V.1}
 \ee
 where $\;{\bf \nabla}U\;$ has components
 \begin{eqnarray}
 \left\{
 \begin{array}{ccc}
 \partial_xU & = & \displaystyle{\frac{\rho_e^2\;J_2}{r^4}\left[\frac{x}{r}\left(\frac{15}{2}
 \;\sin^2 \phi\;-\;\frac{3}{2}\right)\;-\;3\;\sin \phi\;\sin I_p\;\sin h_p\right] }\\
 \\
 \partial_yU & = & \displaystyle{\frac{\rho_e^2\;J_2}{r^4}\left[\frac{y}{r}\left(\frac{15}{2}
 \;\sin^2 \phi\;-\;\frac{3}{2}\right)\;+\;3\;\sin \phi\;\sin I_p\;\cos h_p\right]}\\
 \\
 \partial_zU & = & \displaystyle{\frac{\rho_e^2\;J_2}{r^4}\left[\frac{z}{r}\left(\frac{15}{2}
 \;\sin^2 \phi\;-\;\frac{3}{2}\right)\;-\;3\;\sin \phi\;\cos I_p\right]}~~~.~~~
 \end{array}
 \right.
 \label{V.2}
 \end{eqnarray}
 Here $\;\phi\;$ and $\;{\erbold}\,=(x,y,z)$ denote, correspondingly, the latitude of Deimos
 relative to the Martian equator and the position vector of Deimos related to the Martian
 center of mass; $\,\rho_e\,$ is the Martian equatorial radius; $\,M\,$ and $\,m\,$ stand for
 the masses of Mars and Deimos, respectively. Angles $\;h_p\;$ and $\;I_p\;$ are the longitude
 of the node and the inclination of the planet's equator of date relative to the invariable plane (see Section \ref{subsec:Calcofcomp}). 
 Integration in this, inertial, frame offers the
 obvious advantage of nullifying the inertial forces.

Table 2 gives the initial conditions for our simulation, expressed
in terms of the Keplerian orbital elements. Table 3 presents these
initial conditions in a more practical form, i.e., in terms of the
Cartesian positions and velocities corresponding to the said
elements. A transition from the Keplerian elements to these
Cartesian positions and velocities is a two-step process. First, we
take orbital elements defined in a frame associated with the Martian
equator of date (i.e., in a frame co-precessing but not co-rotating
with the planet) and transform them into Cartesian coordinates and
velocities defined in that same frame. Then, by two successive
rotations of angles $\;-\;I_p\;$ and $\;-\;h_p\;$, we transform them
into Cartesian coordinates and velocities related to the invariable plane.
These initial positions and velocities were used to
begin the integration.

At each step of integration of (\ref{V.1}), the same two rotations
are performed on the components $\;{\bf\nabla} U\;$ given by
(\ref{V.2}). As mentioned above, to afford the absence of inertial
forces on the right-hand side of (\ref{V.1}) one must write down and
integrate (\ref{V.1}) in the inertial frame. Since the analytical
expressions (\ref{V.2}) for $\;{\bf\nabla} U\;$ contain the latitude
$\;\phi\;$, they are valid in the co-precessing coordinate system
and, therefore, need to be transformed to the inertial frame at each
step. To carry out the transformation, one needs to know, at each
step, the relative orientation of the Martian polar axis and the
inertial coordinate system. The orientation is given by the afore
mentioned Colombo model. This is how our second equation, the one of
Colombo, comes into play:
 \begin{eqnarray}
 \frac{d{\bf k}}{dt}&=&\alpha({\bf \hat{k}\cdot \hat{n}})({\bf
 \hat{k}\times \hat{n}})\;\;\;.
 \label{V.3}
 \end{eqnarray}
All in all, we have to integrate the system (\ref{V.1} - \ref{V.3}).
Table 4 gives the initial conditions used for integrating
(\ref{V.3}), while Table 5 gives the numerical values used for the
parameters involved. It is worth noting that the values of Table 4
were calculated based on Ward (1974).

 The software used for numerical integration of the system (\ref{V.1} - \ref{V.3}) is called
 NOE (Numerical Orbit and Ephemerides), and is largely based on the ideas and methods
developed in Lainey, Duriez \& Vienne (2004). This numerical tool
was created at the Royal Observatory of Belgium mainly for
computations of the natural satellite ephemerides. It is an $N$-body
code, which incorporates highly sensitive modelling and can generate
partial derivatives. The latter are needed when one wants to fit the
initial positions, velocities, and other parameters to the
observation data. To save the computer time, an optimised force
subroutine was built into the code, specifically for integrating the
above equations. This appliance, based on the RA15 integrator
offered by Everhart (1985), was chosen for its speed and accuracy.
During the integration, a variable step size with an initial value
of $\,0.04\,$ day was used. To control the numerical error, back and
forth integrations were performed. In particular, we carried out a
trial simulation consisting of a thousand-year forward and a
subsequent thousand-year back integration. The satellite
displacement due to the error accumulated through this trial was
constrained to $\,150\,$ meters. Most of this $\,150\,$ meter
difference comes from a numerical drift of the longitude, while the
numerical errors in the computation of the semi-major axis, the
eccentricity, and the inclination were much lower. These errors were
reduced for this trial simulation to only $\,10^{-5}\,km\,$,
$\,10^{-10}\,$, and $10^{-10}$ degree, respectively. This provided
us with a high confidence in our subsequent numerical results.

As a complement to the said back-and-forth check, the
energy-conservation criterium was used to deduce, in the first
approximation, an optimal initial step-size value and to figure
out the numerical error proliferation. (It is for this
energy-conservation test that we introduced a non-zero mass for
Deimos. Its value was taken from Smith, Lemoine \& Zuber (1995).)
Applicability of this criterium is justified by the fact that the
numerical errors are induced mostly by the fast orbital motion of
the satellite.\footnote{Although the planet-satellite system is
subject to an external influence (the solar torque acting on the
planet), over short time scales this system can be assumed closed.
In order to check the integrator efficiency and to determine an
optimal initial step size, we carried out auxiliary integrations
of (\ref{V.1}) - (\ref{V.2}), with the Colombo equation
(\ref{V.3}) neglected and with the energy presumed to conserve.
These several-thousand-year-long trial integrations, with the
energy-conservation criterium applied, led us to the conclusion
that our integrator remained steady over long time scales and that
the initial step of 0.04 day was optimal. Then this initial step
size was employed in our integration of the full system
(\ref{V.1}) - (\ref{V.3}).}

\begin{table}[htbp]
\begin{center}
\begin{tabular}{lc}
\hline
parameters&numerical values\\
\hline
$a$ & 23459 km\\
\hline
$e$& 0.0005\\
\hline
$i$& 0.5 deg and 89 deg\\
\hline
$\Omega$& 10 deg\\
\hline
$\omega$& 5 deg\\
\hline
$M$ & 0 deg.\\
\hline
\end{tabular}
\end{center}
 \caption{The orbital elements values taken as initial conditions for our simulations.}
 \label{table1}
\end{table}

\begin{table}[htbp]
\begin{center}
\begin{tabular}{lccc}
\hline
satellite &$x$&$y$&$z$\\
\hline
position ${\rm km}$ ($i=0.5$) &  22648.3376439  &      6068.52353055   &     17.8332361962 \\
\hline
velocity  ${\rm km/s}$ ($i=0.5$) & -0.349882011871   &     1.30576017694   &    1.175229063323$\;\times\; 10^{-2}$ \\
\hline
position ${\rm km}$ ($i=89$) &   22996.9921622   &     4091.20549954    &    2043.25303109   \\
\hline
 velocity ${\rm km/s}$ ($i=89$) &  -0.120115009144  &     2.686751629968$\;\times\; 10^{-3}$ &  1.34652528539 \\
\hline
\end{tabular}
\end{center}
 \caption{The initial positions and velocities used for Deimos. The first two rows correspond to the low-inclination case (0.5
 degree); the last two rows correspond to the high-inclination case (89 degrees).}
\end{table}

\begin{table}[htbp]
\begin{center}
\begin{tabular}{lc}
\hline
parameters&numerical values\\
\hline
$I_p(t_0)$&$25.25797549$ deg\\
\hline
$h_p(t_0)$&$332.6841708$ deg\\
\hline
\end{tabular}
\end{center}
 \caption{Initial conditions used for the integration of Eq.~(\ref{V.3}). These values were calculated based on Ward (1974).}
 \label{table3}
\end{table}

\begin{table}[htbp]
\begin{center}
\begin{tabular}{lc}
\hline
parameters&numerical values\\
\hline
Martian mass (GM)&$42830\ {\rm km^3s^{-2}}$\\
\hline
$J_2$&$1960.45\ \times   10^{-6}$\\
\hline
Equatorial radius& $3397\ {\rm km}$\\
\hline
Deimos mass&$0.091\ \times  10^{-3}\ {\rm km^3s^{-2}}$\\
\hline
$\alpha$&  $3.9735\ \times  10^{-5}$ {\rm rad/yr}\\
\hline
\end{tabular}
\end{center}
\caption{Parameter values used in our simulations.}
\end{table}

~\\

 \subsection{Integration by the semianalytical approach}

 The theory of satellite-orbit evolution, based on the planetary equations whose
 right-hand sides combine those of (\ref{1} - \ref{5}) and (\ref{3body3}), is
 semianalytical. This means that these equations for the elements' secular parts
 are derived analytically, but their integration is to be performed
 numerically. This integration was carried out using an $8^{th}$-order
 Runge-Kutta scheme with relative and absolute tolerances of
 $\;10^{-12}\;$. Kilograms, years, and kilometers were taken as
 the mass, time, and length units, correspondingly.

 \subsubsection{\textbf{Technicalities}}

 To integrate the planetary equations, one should know, at each time step,
 the values of $\;\dot I_p\;$ and $\;\dot h_p\;$, which are the time derivatives of the
 inclination and of the longitude of the node of the equator of date with respect to the
 equator of epoch. These derivatives will be rendered by the Colombo equation
 (\ref{20}), after formulae (\ref{18} - \ref{19}) get inserted therein:
 \begin{eqnarray}
  \nonumber
 \dot I_p\;=\;-\;\alpha\;\left(\;{q}^{2}\;\sin { I_p}\;\sin { h_p}\;\cos {h_p}\;-\;q\;p\;\sin
 {I_p}\;+\;2\;p\;q\;\sin {I_p}\;\cos^{2}{h_p}
   \right.
   \nonumber \\
   \nonumber\\
   \nonumber
   -  \;{p}^{2}\;\sin {I_p} \; \cos {h_p}\;\sin {h_p}\;+\;q\;\sqrt
 {1\;-\;{q}^{2}\;-\;{p}^{2}\;}\;\cos {I_p}\;\cos {h_p}~~~~~~~
 \nonumber\\
 \nonumber\\
 \left.
   -  \;p \;\sqrt{1\;-\;{q}^{2}\;-\;{p}^{2}\;} \;\cos {I_p} \;\sin {h_p}\;\right)\;\;\;,
   ~~~~~~~~~~~~~~~~~~~~~~~~~~~~~~~~~~
 \label{38}
 \end{eqnarray}
 and
 \begin{eqnarray}
 \nonumber
 {\dot{h}}_ p \,=
 \,-\,\alpha\,\left\{\left[\left(p - 2p\,\cos^{2} {I_p} \right)\,\cos h_p
 +\,\frac{\left(\,-\,q + 2 q\,\cos^{2}{I_p}\right)\,\cos^{2} h_p -\,2
 q\,\cos^{2}I_p + q}{\sin h_p} \right]\,\frac{\sqrt{1 - p^2 - q^2}
 }{\sin {I_p}}\right.
 \\
 \nonumber\\
 \nonumber\\
 \nonumber
 +\;\left({q}^{2}\,-\,{p}^{2} \right) \;\,\cos {I_p}\;\,\cos^{2} {h_p}\,+\,
 \left(\,-\,{p}^{2}\,-\,2\,{q}^{2}\,+\,1\right)\;\,\cos{I_p}~~~~~~~~~~~~~~~~
 ~~~~~~~~~~~~~~~~~~~~~~~~~~~~~~~~
 \ea
 \ba
 \left.
 +~~\frac{2\,q\,p~\,\cos {I_p}~\,\cos^{3} {h_p}\,-\,2\,p\,q~\,\cos {h_p}~\,
 \cos
 {I_p}}{\sin{h_p}}~\right\}~~.~~~~~~~~~~~~~~~~~~~~~~~~~~~~~~~~~~~~~~\,~~~~~
 \label{39}
 \end{eqnarray}
 Equations (\ref{38}) and (\ref{39}) are then integrated (with the initial
 conditions $\,I_p(t_0)\,$ and $\,h_p(t_0)\,$ borrowed from Table
 \ref{table3}) simultaneously with the planetary equations (\ref{1} -
 \ref{5}). Through formulae (\ref{6}), the above expressions for $\;{\dot{I}
 }_p\;$ and $\;{\dot{h}}_p\;$ yield the expressions for the components of $
 \;\mubold\;$. As can be seen from (\ref{12} - \ref{14}), integration of
 the planetary equations also requires the knowledge of the derivatives $\,\dot{
 \mu}_1\,,\;\dot{\mu}_2\,$, and $\,\dot{\mu}_3\,$ at each integration step.
 These can be readily obtained by differentiating (\ref{6}). The resulting
 closed-formed expressions for $\,\dot{\mu}_1\,,\,\dot{\mu}_2\,,\,\dot{\mu
 }_3\,$ are listed in Appendix A. The final step required for numerical
 integration of the planetary equations is substitution of formulae
 (\ref{9} - \ref{11}), with the initial conditions from Table \ref{table1}.

 \subsubsection{\textbf{The plots and their interpretation}}

\begin{center}
\begin{figure}[h]
  \includegraphics[width=6.0in]{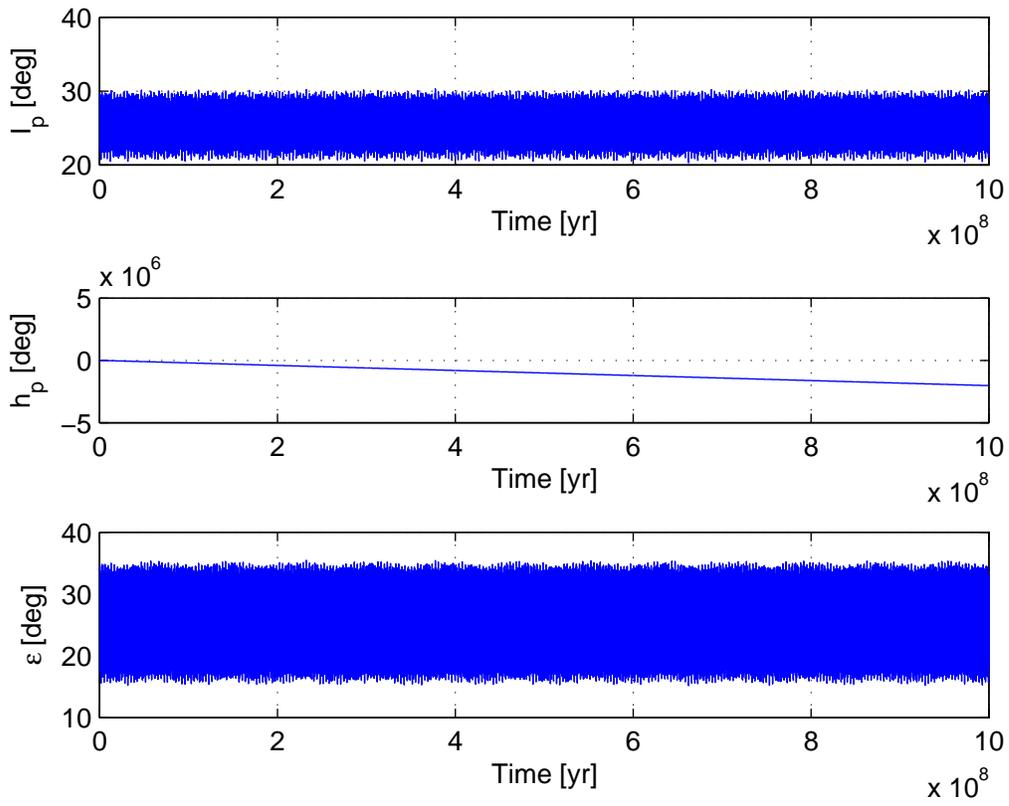}\\
  \caption{Evolution of the Martian inclination, the longitude of the node of the
  equator of date relative to that of epoch, and the obliquity over
  1 Byr, obtained in the Colombo approximation.}
  \label{fig1_paper}
\end{figure}
\end{center}

\begin{center}
\begin{figure}[h]
  \includegraphics[width=6.0in]{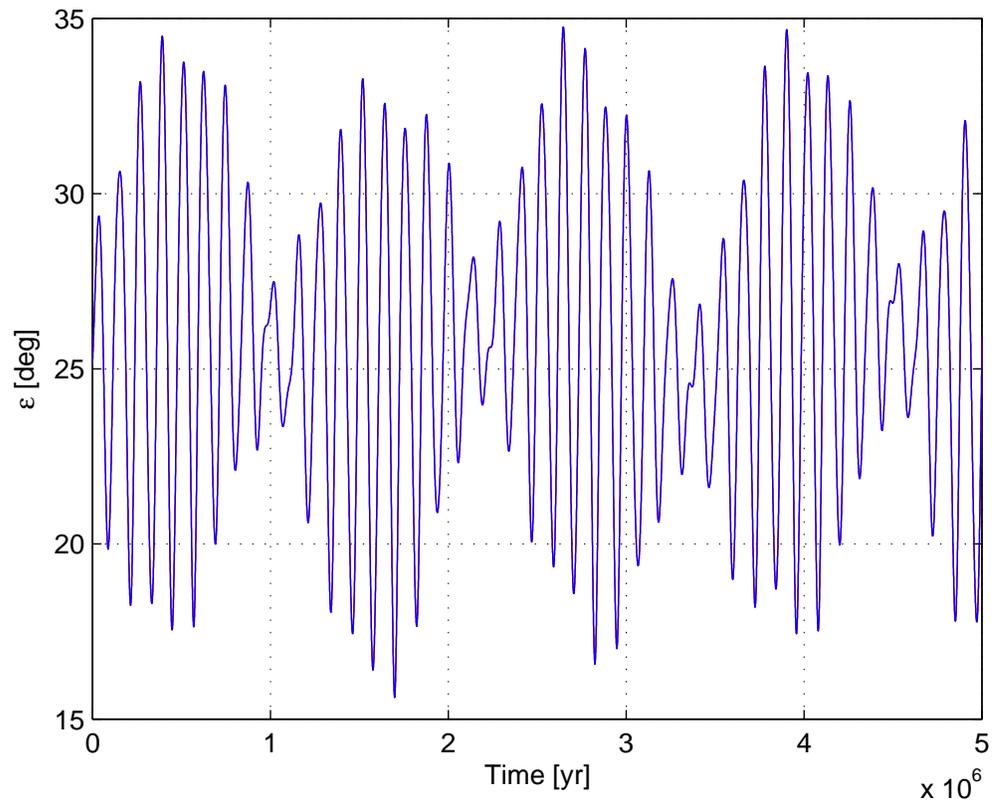}\\
  \caption{Evolution of the the Martian obliquity for 5 Myr,
  calculated through formula (\ref{oblo}). This curve
  closely matches the result
  of Ward (1974).}
  \label{fig2_paper}
\end{figure}
\end{center}

 Fig.~\ref{fig1_paper} depicts the history of the planetary equator, in the
 Colombo approximation, over 1 Byr. The inclination exhibits
 long-periodic oscillations bounded within the range of $\;20.3\,\deg\,\le\,
 I_p\,\le\,30.3\,\deg\,$, while the node regresses at a rate of $\;\dot{h}_p
 \,=\,0.00202\,\deg/\yr\,$. The obliquity, $\epsilon$, varies in the
 range $\;15.2\,\deg\,\le\,\epsilon\,\le\,35.5\,\deg\,$. A magnified
 view of the obliquity for 5 Myr is shown in Fig.~\ref{fig2_paper}.
 The time history of the obliquity closely matches the results
 reported by Ward (1974).

\begin{center}
\begin{figure}[h]
\begin{tabular}{c}
  \includegraphics[width=6.0in]{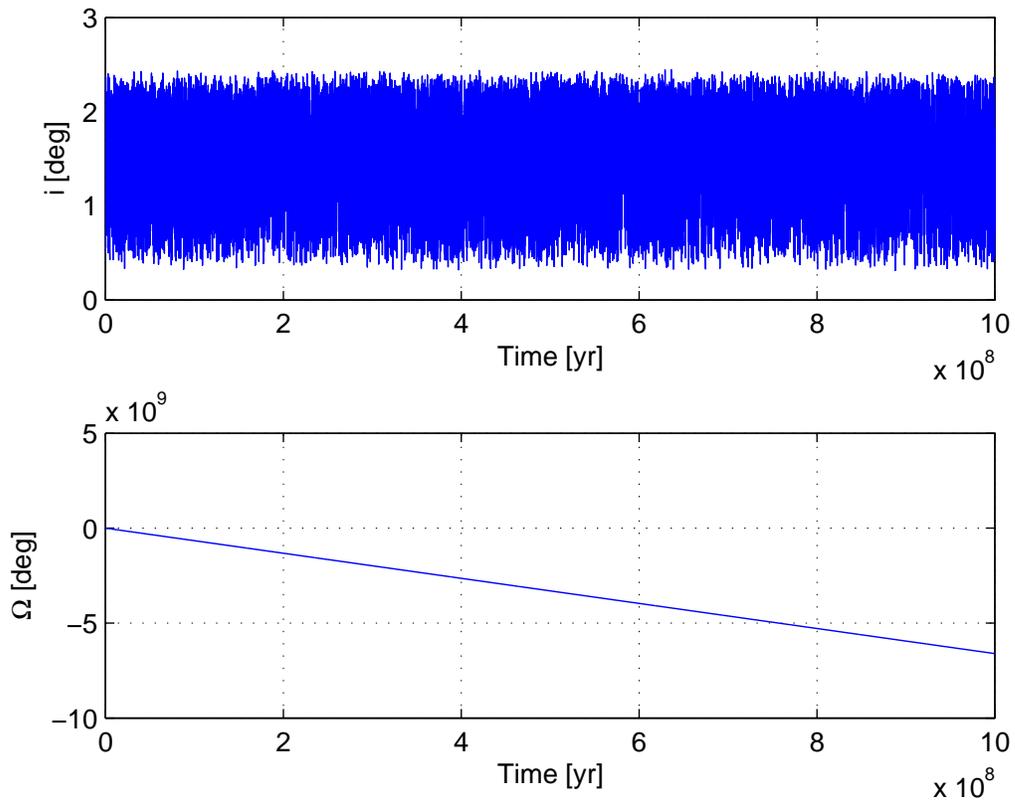}\\
\end{tabular}
  \caption{Evolution of the inclination (initially set to 0.5 degree)
  and of the longitude of the node of Deimos over 1 Byr. The plot,
  obtained by integration of the semianalytical model, exhibits
  inclination locking and a uniform regression of the node.}
  \label{fig3_paper}
\end{figure}
\end{center}

\begin{center}
\begin{figure}[h]
\begin{tabular}{c}
  \includegraphics[width=6.0in]{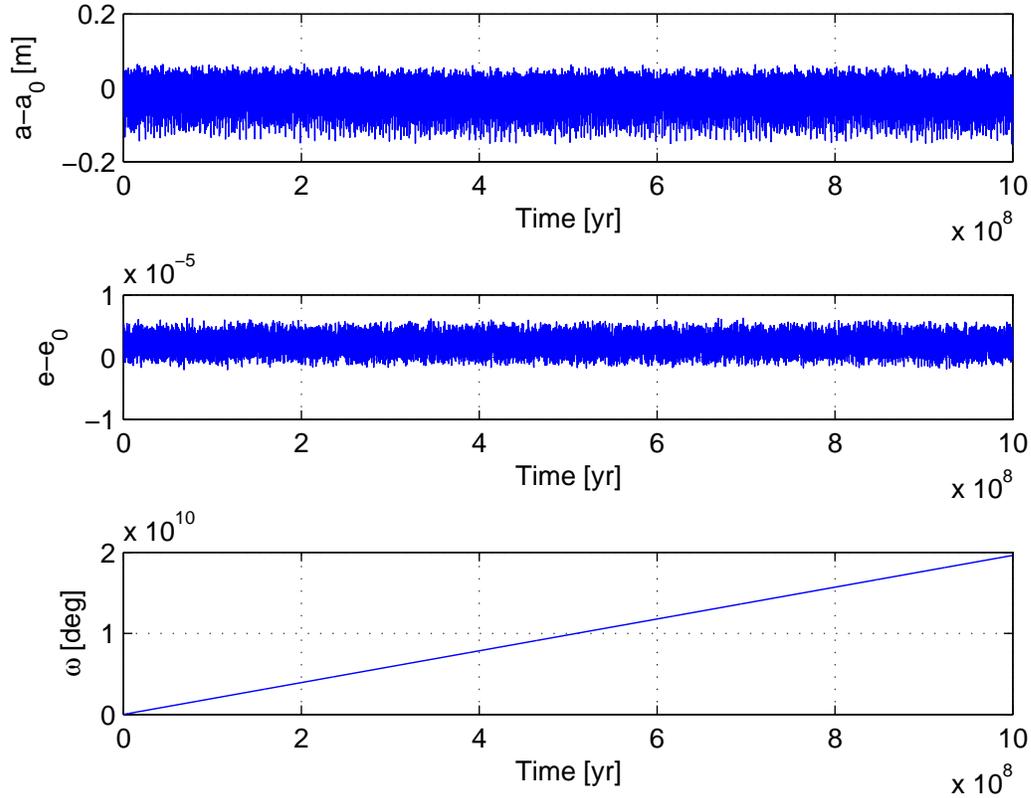}\\
\end{tabular}
 \caption{Evolution of the semimajor axis, eccentricity, and argument of periapsis of
 Deimos over 1 Byr. (The inclination was initially set to 0.5 degree.) Both the semimajor
 axis and eccentricity exhibit quasiperiodic motion about their initial values. (The
 variations of the semimajor axis and eccentricity are so small that it is more convenient to plot
 $\;a-a_0\;$ and $\;e-e_0\;$.) The plots were obtained by integration of the
 semianalytical model.}
 \label{fig4_paper}
\end{figure}
\end{center}

\begin{center}
 \begin{figure}[h]
 \includegraphics[width=5.5in]{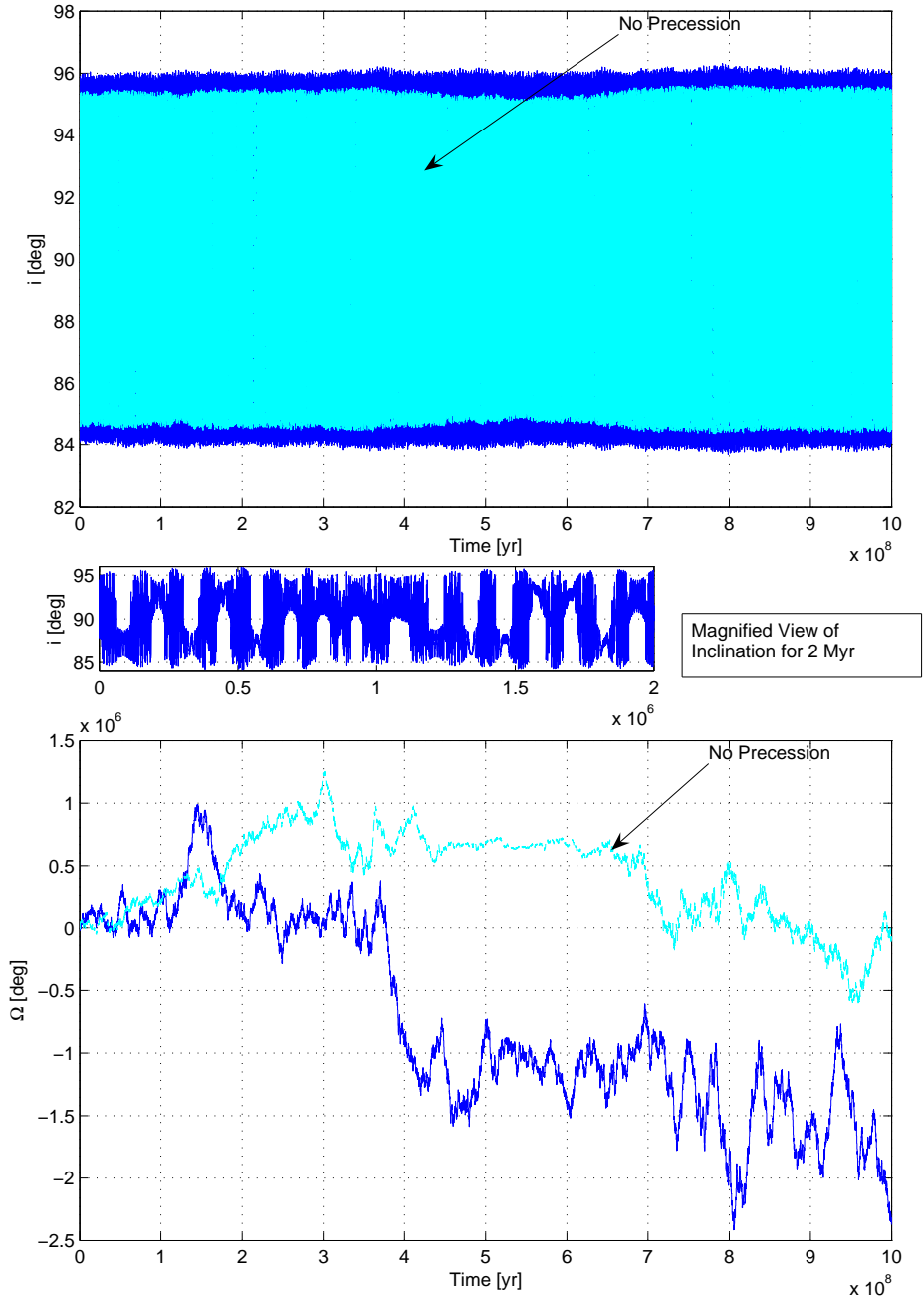}\\
 \caption{Evolution of the inclination and of the longitude of the node of Deimos over
 1 Byr. (The inclination was initially set to 89 degrees.) The plot, obtained by
 integration of the semianalytical model, exhibits inclination variation in the range $\pm 5$ deg and a
 chaotic
 evolution of the node; if precession is neglected, the inclination oscillations are smaller in magnitude. The chaotic nature of the inclination variation is
 referred to as ``crankshaft chaos".}
 \label{fig5_paper}
 \end{figure}
 \end{center}

\begin{center}
\begin{figure}[h]
  \includegraphics[width=6.0in]{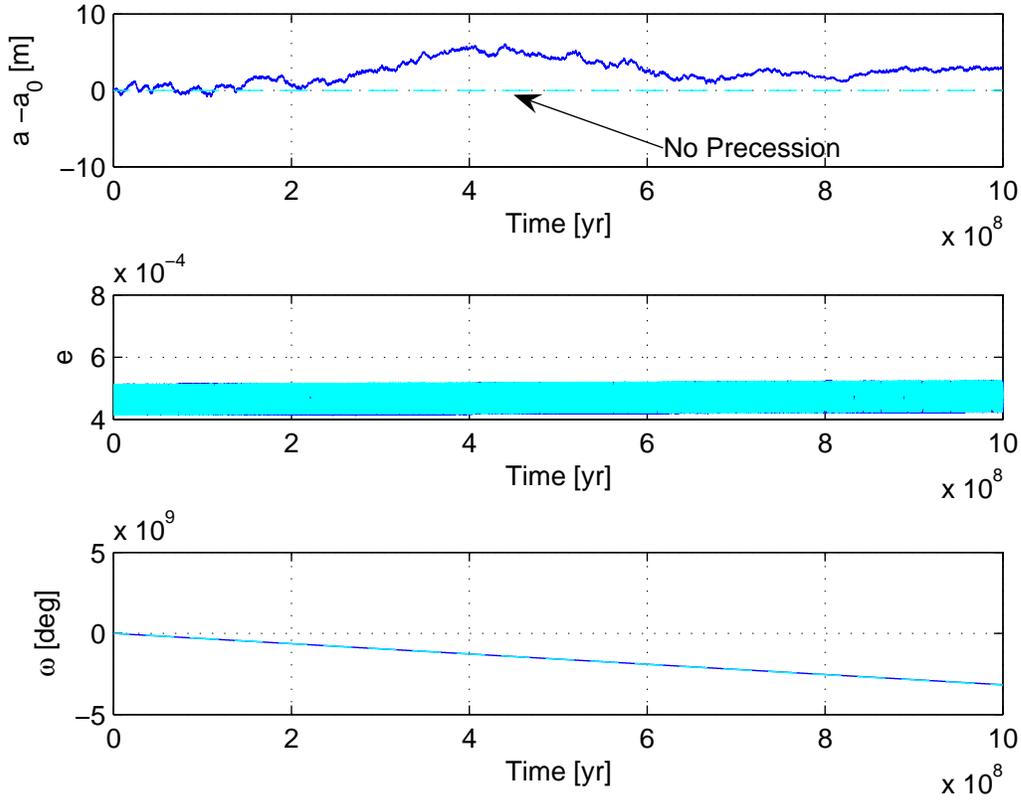}\\
  \caption{Evolution of the semi-major axis, eccentricity, and argument of
  periapsis of Deimos over 1 Byr. (The inclination was initially set to 89 degrees.)
  The semi-major axis exhibits chaotic behavior and the eccentricity exhibits long-periodic motion about the initial
  value. (The variations of the semimajor axis are so small that it is more convenient to
  plot $\;a-a_o\;$ rather than $\;a\;$.) The plots were obtained by integration of the
  semianalytical model.}
  \label{fig6_paper}
\end{figure}
\end{center}

 Integration of our semianalytical model gives plots depicted in Figures
 \ref{fig3_paper} and \ref{fig4_paper}, for a low initial inclination, and in
 Figures \ref{fig5_paper}-\ref{fig6_paper}, for a high initial inclination.
 From Fig.~\ref{fig3_paper} we see that the variable equinoctial precession
 does not inflict
 considerable changes upon the satellite's inclination relative to the precessing
 equator of date. The orbit inclination remains bounded within the region
 $\,0.3\,\deg\,\le\,\inc\,\le\,2.5\,\deg\,$. This means that the
 ``Goldreich lock" (inclination ``locking," predicted by the Goldreich
 (1965) model for small inclinations and for uniform equinoctial precession)
 works also for nonuniform Colombo precession of the equator.

However, this observation no longer holds for large initial
inclinations. In the $\inc_0=89$ deg case, it is clearly seen that
the node is greatly affected by the presence of the equinoctial
precession. The equinoctial precession also affects the magnitude of
the inclination variations. This case exhibits chaotic dynamics that
are sensitive to any additional perturbing inputs.
Fig.~\ref{fig5_paper} shows that without precession the inclination
of Deimos' orbit varies within the range of
$\,84.5\,\deg\,\le\,\inc\,\le\,95\,\deg\,$. With the precession
included, the inclination gains about one degree in amplitude,
varying in the range $\,83.5\,\deg\,\le\,\inc\,\le\,96\,\deg\,$.
This effect is accentuated when examining a magnification of the
inclination over a 5 Myr span, as shown in Fig.~\ref{fig5_paper}.
The chaotic nature of the inclination is clearly seen. The irregular
dynamics is characterised by chaotic switches between the maximum
and minimum inclination values, a phenomenon referred to as
``crankshaft", to be further discussed in the sequel.

 Both in the near-equatorial case (as in Fig.~\ref{fig4_paper}) and the
 near-polar case (as in Fig.~\ref{fig6_paper}),
 variations of the semimajor axis are of order $\;10^{-6}\,$\%$\,$. This
 smallness is in compliance with formula (44) in Efroimsky (2006a,b),
 according to which the changes of $\,a\,$ generated by the variations of
 the equinoctial precession rate are extremely small. (Formula (\ref{vavila})
 from subsection 2.4 above tells us that the direct pull of the Sun
 exerts no influence upon $\,a\,$ at all.)

 Similarly, in both cases (Fig.~\ref{fig4_paper} and Fig.~\ref{fig6_paper})
 the variations of eccentricity, remain small, about $\;10^{-3}\,$\%$\,$.
 While formula (44) in Efroimsky (2006a,b) promises to $\,e\,$ only tiny
 variations due to precession, our formula (\ref{gloria}) from subsection
 2.4 above gives to $\,e\,$ a slightly higher variation rate, rate that
 still remains insufficient to raise the quasiperiodic changes in $\,e\,$
 above a fraction of percent.

 The plots in Figs.~\ref{fig4_paper} and \ref{fig6_paper} depict also the time
 evolution of $\,\omega\,$. The line of apsides steadily regresses in the
 near-polar case and steadily advances in the near-equatorial case.

\begin{figure}
\begin{center}
 \includegraphics[width=12cm]{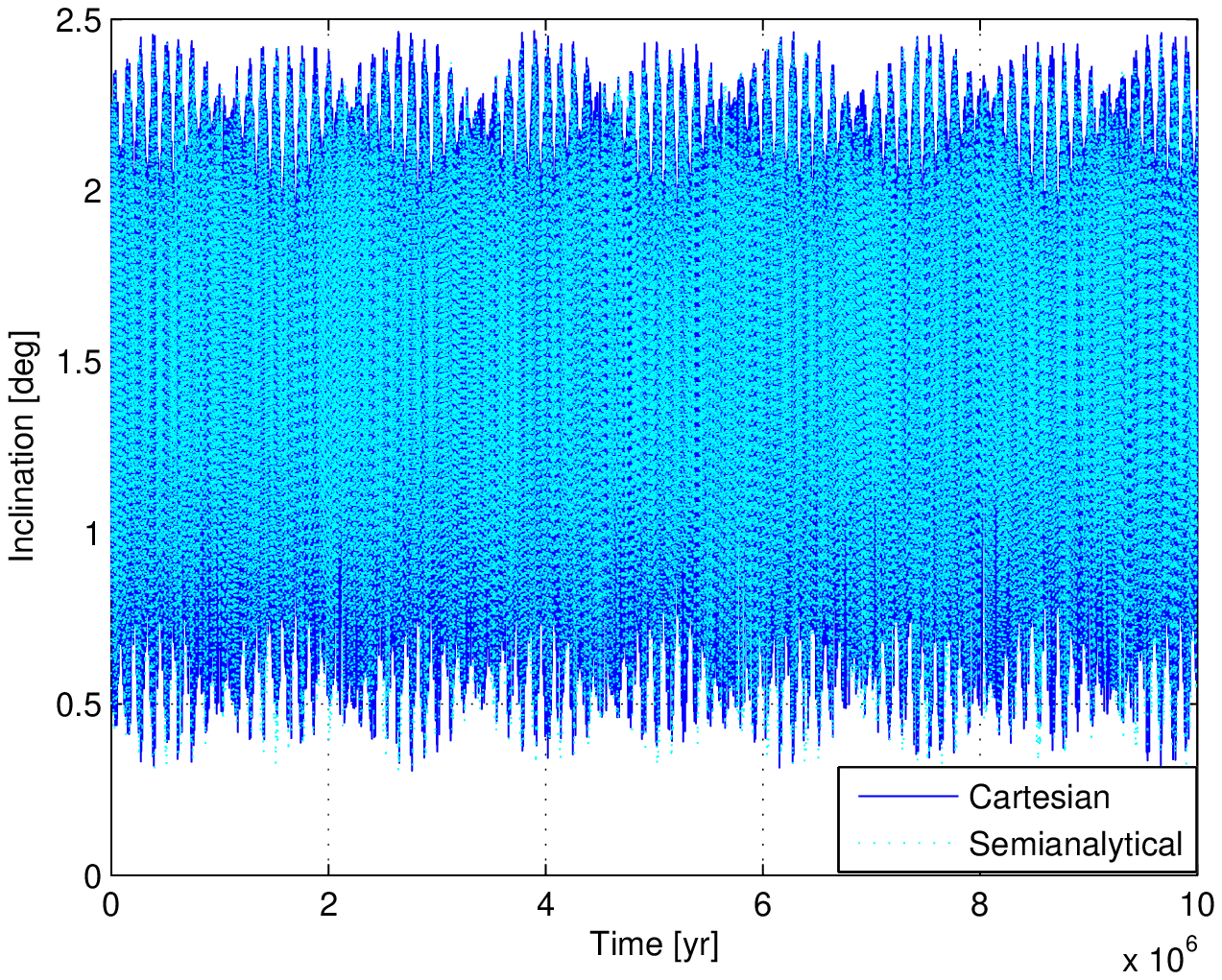}
 \includegraphics[width=12cm]{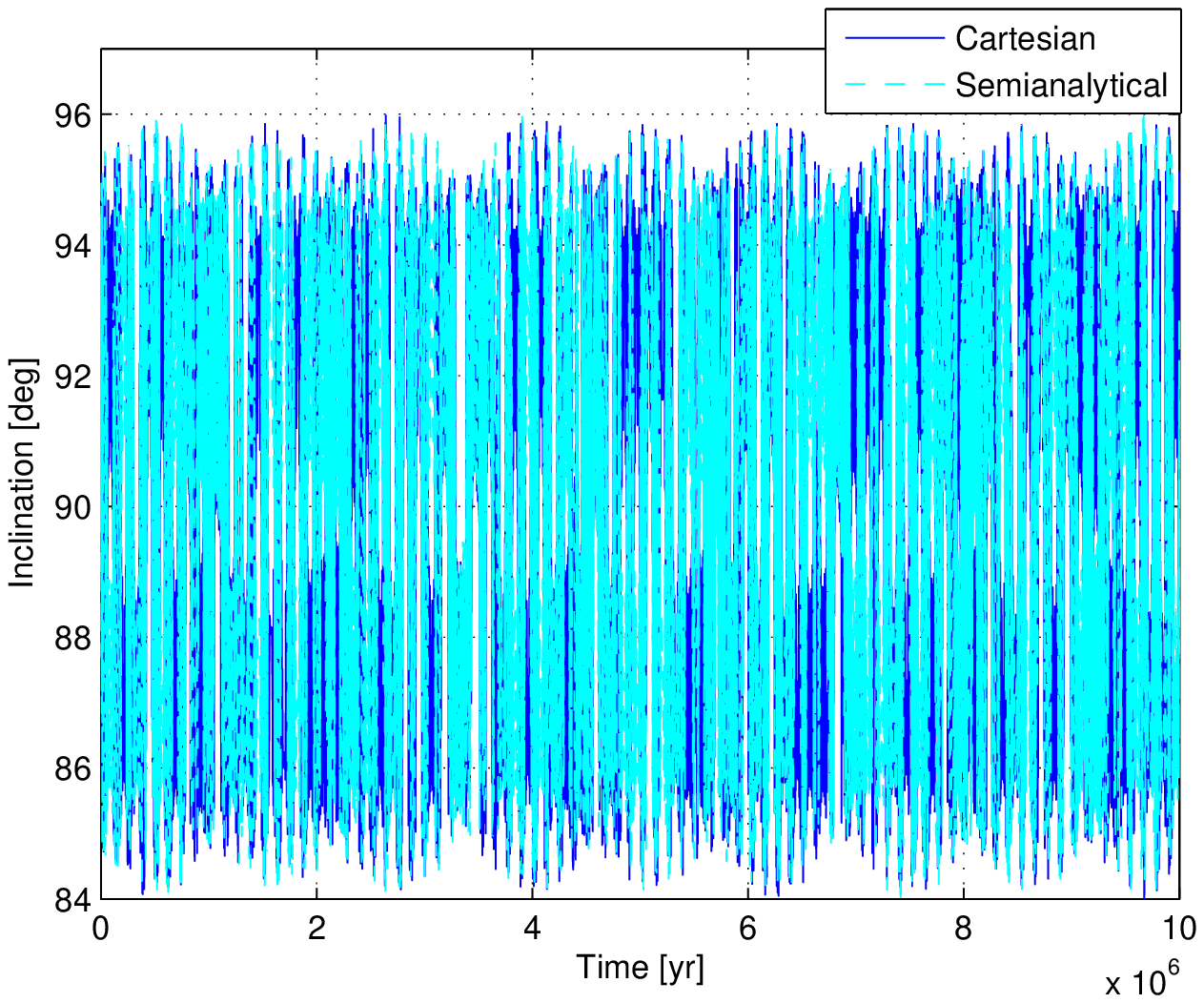}
 \caption{Comparison
of
  the semi-analytical model to a purely numerical integration. The plot
  shows the inclination as predicted by the two models. The top plot shows the case of
  the initial inclination $\,\inc_0\,=\,0.5\,\deg$. In this case,
  the standard deviation of the inclination in the semianalytical
  and inertial models differs by 0.175\%, and the mean value by
  0.77\%. The bottom plot shows the case of
  the initial inclination $\,\inc_0\,=\,89\,\deg$. In this case,
  the differences between the semianalytical and the numerical model
  amount to 0.4\% in standard deviation and 0.18\% in the mean value.
  \label{fig8_paper}}
\end{center}
\end{figure}

 To examine the precision of our semianalytical model, we have compared the
 results of its integration with the results stemming from a purely numerical
 simulation performed in terms of inertial Cartesian coordinates and
 velocities (see subsection 3.1). The necessity for this check was dictated,
 mainly, by the fact that within the semianalytical model the short-period
 terms are averaged out,\footnote{~We remind that in equations (\ref{1} -
 \ref{5}) the exact $\;\mubold$-dependent terms are substituted with their
 orbital averages (\ref{12} - \ref{14}).
 } while the straightforward numerical integration of
 (\ref{V.1} - \ref{V.2}) neglects nothing. We carried out comparison of
 the two methods over 10 Myr only. The outcomes, both for $\,\inc_0\,=\,0.5
 \,$ degrees and $\,\inc_0\,=\,89\,\deg\,$, were in a good agreement.
 As an example, the top plot in Fig.~\ref{fig8_paper} shows the comparison
 of the inclination evolution calculated by the semianalytical and purely
 numerical methods over 10 Myr in the case of $\,\inc_0\,=\,0.5\,\deg\,$.
 The bottom plot in Fig.~\ref{fig8_paper} depicts a similar comparison for
 the case of $\,\inc_0\,=\,89\,\deg$. Since the inclination exhibits chaotic
 behaviour, comparing the semianalytical and purely numerical calculations
 point-by-point (i.~e., computing the differences between these two signals)
 would not be useful. Therefore, we chose to compare the mean and standard
 deviation (STD) of the inclination, in the semianalytical and purely
 numerical simulations.
 We also compared the extremum values. The results of this comparison are
 summarised by Table \ref{std}. The table quantifies the agreement between
 the models depicted by Fig.~\ref{fig8_paper}. It clearly demonstrates
 that the two simulations agree up to fractions of a percent.


 All in all, the outcome of our computations is two-fold. First, we have
 made sure that the semianalytical model perfectly describes the dynamics
 over time scales of, at least, dozens of millions of years. Stated
 differently, the short-period terms and the terms of order $\,O(\mubold^2)
 \,$ play no role over these time spans. Second, we have made sure that the
 ``Goldreich lock" initially derived for very low inclinations and for
 uniform equinoctial precession, works well also for variable precession,
 though for low initial inclinations only.

 \begin{table}[h]
 \begin{center}
 \begin{tabular}{|c|c|c|c|c|c|}
 \hline
  $\inc_0$ [deg]& Model & STD [deg] & Mean [deg] & Max. Value [deg] & Min. Value [deg]\\
  \hline\hline
  0.5 & Semianalytical         & 0.60 &  1.519 & 2.45 & 0.3063 \\
      & Cartesian              & 0.601 & 1.53 & 2.465  & 0.3056 \\
      \hline
  89 & Semianalytical          & 3.10 & 90.085 & 95.9713 &84.027\\
     & Cartesian               & 3.09 & 89.92 & 95.9769 & 84.0054\\
 \hline
 \end{tabular}
 \caption{Statistical properties of the inclination. Comparison of the results
 of the semianalytical and the purely numerical computation.}
 \label{std}
 \end{center}
 \end{table}
 \subsubsection{\textbf{Looking for trouble}}

 A natural question arises as to whether the considered examples are
 representative. One may enquire if, perhaps, there still exists a
 combination of the initial conditions yielding noticeable variations of the
 satellite orbit inclination during the primary's variable
 equinoctial precession over vast spans of time. To answer this question, we
 should scan through all the possible combinations of initial conditions, to
 identify a particular combination that would entail a maximal inclination
 excursion relative to the initial inclination. Stated more formally, we
 should seek a set of initial conditions $\;\{\,h_{p0}^{\star}\,,\;I_{p0}^{
 \star}\,,\;\inc_0^{\star}\,,\;\Omega_0^{\star}\,,\;\omega_0^{\star}\,\}\;$
 maximising the objective function $\;|\,\inc(t)\,-\,\inc_0\,|\;$:
 \ba
 \{h_{p0}^{\star},\, I_{p0}^{\star} ,\, i_0^{\star},\, \Omega_0^{\star},\,
 \omega_0^{\star}\}=\underset{i_0,\Omega_0,\omega_0}{\underset{t,h_{p0},I_{
 p0}}{\arg\max}} |i(t)-i_0 |
 \label{41}
 \label{opt}
 \ea
 This problem belongs to the realm of optimisation theory. The optimisation
 space is constituted by the entire multitude of the permissible initial
 conditions. The sought after combination of initial conditions will be
 called $\,${\it{the optimal set.}}

 In this situation, the traditional optimisation schemes (such as the
 gradient search or the simplex method) may fail due to the rich dynamical
 structure of our problem -- these methods may lead us to a local extremum
 only.
 Thus, the search needs to be global. It can be carried out by means of
 Genetic Algorithms (GA's) \cite{goldberg, gurfil1, gurfil2, gurfil3}. These
 are wont to supersede the traditional optimisation procedures in the
 following aspects. First, instead of directly dealing with the parameters,
 the GA's employ codings (usually, binary) of the parameter set (``strings,"
 in the GA terminology). Second, instead of addressing a single point of the
 optimisation space, the GAs perform a search inside a population of the initial
 conditions. Third, instead of processing derivatives or whatever other
 auxiliary information, the GAs use only objective-function evaluations
 (``fitness evaluations"). Fourth, instead of deterministic rules to
 reiterate, the GAs rely upon probabilistic transition rules. Additional
 details on the particular GA mechanism used herein can be found in Appendix
 B.

\begin{center}
\begin{figure}[h]
  \includegraphics[width=5.0in]{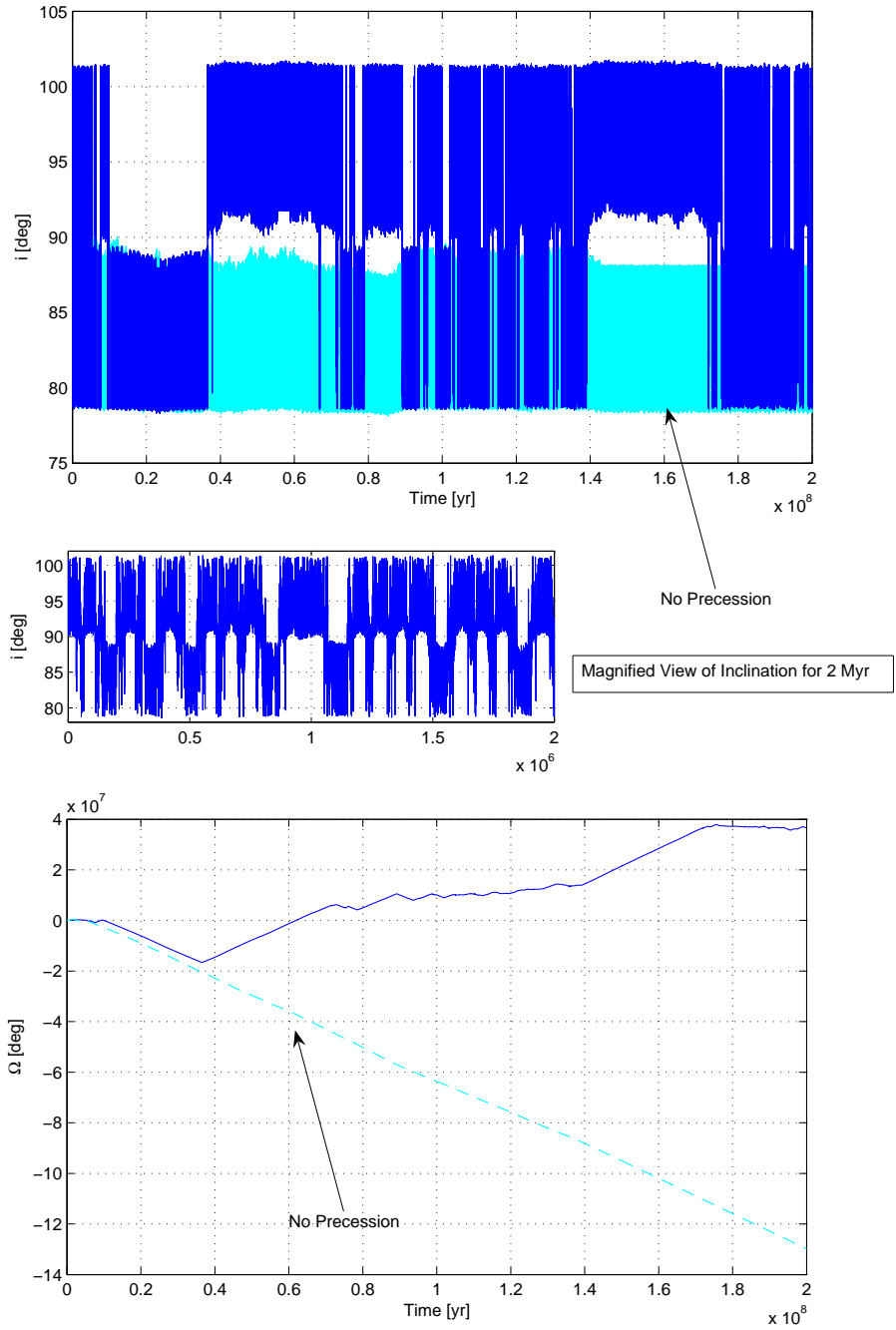}\\
  \caption{Evolution of the inclination and longitude of the node, for
  the initial conditions that entail maximal variations of the
  orbit inclination. The plot, obtained by integration of the semianalytical
  model, demonstrates that the inclination, initially set at 100.5
  deg, plunges to less than 80 deg and then returns to its initial
  value. The switches between maximum and minimum values
  exhibit ``crankshaft" chaos. Without precession, the
  inclination magnitude is much smaller, and the node regresses uniformly.}
  \label{fig9_paper}
\end{figure}
\end{center}

 A GA optimisation was implemented using the parameter values given in Table
 \ref{tab:ga}.
 \begin{table}[htbp]
 \begin{center}
 \begin{tabular}{lc}
 \hline
 parameters&numerical values\\
 \hline
 Population size  &$30$\\
 \hline
 Number of generations &$100$\\
 \hline
 String length& $16 \, {\rm bit}$\\
 \hline
 Probability of crossover& $0.99$\\
 \hline Probability of mutation&  $0.02$  \\ \hline
 \end{tabular}
 \end{center}
 \caption{Parameter values used for the GA optimization}
 \label{tab:ga}
 \end{table}
 The search for the inclination-maximising initial conditions resulted in
 the following set:
 \ba
 \label{optimalIC}
 I_{p0}^{\star}= 72.5\,\deg\,,\,\; h_{p0}^{\star} = 211.324\,\deg,\,\;
 \inc_0^{\star} =  100.543\,\deg\,,\,\;\Omega_0^{\star} = 111.538\,\deg\,,
 \,\;\omega_0^{\star}=234.913\,\deg ~~~.~~~
 \label{42}
 \ea
 Thus, the initial orbit is \emph{retrograde} and, not surprisingly, \emph{
 near-polar}. The
 resulting time histories for a 0.2~Byr integration are depicted in
 Fig.~\ref{fig9_paper}, for $\,\inc\,$ and $\,\Omega\,$. In both cases the
 inclination amplitude is relatively large: The inclination varies within the range
 of $79 \textrm{ deg}< \inc  < 102 \textrm{ deg}$.

 This example clearly shows that the equinoctial precession is an important
 effect for evolution of satellite orbits. As shown in the upper pane of Fig.~\ref{fig9_paper}, had we
 neglected the precession, the magnitude of the oscillations would
 be about twice smaller. The inclusion of the precession in
 the model qualitatively modifies the behavior, inducing large-magnitude chaotic
 variations of the inclination, a phenomenon that cannot be detected
 without including the precession alongside the oblateness and solar
 gravity.


 \section{Comparison of the semianalytical results\\
  with those rendered by Goldreich's model}

\begin{center}
\begin{figure}[h]
  \includegraphics[width=6.0in]{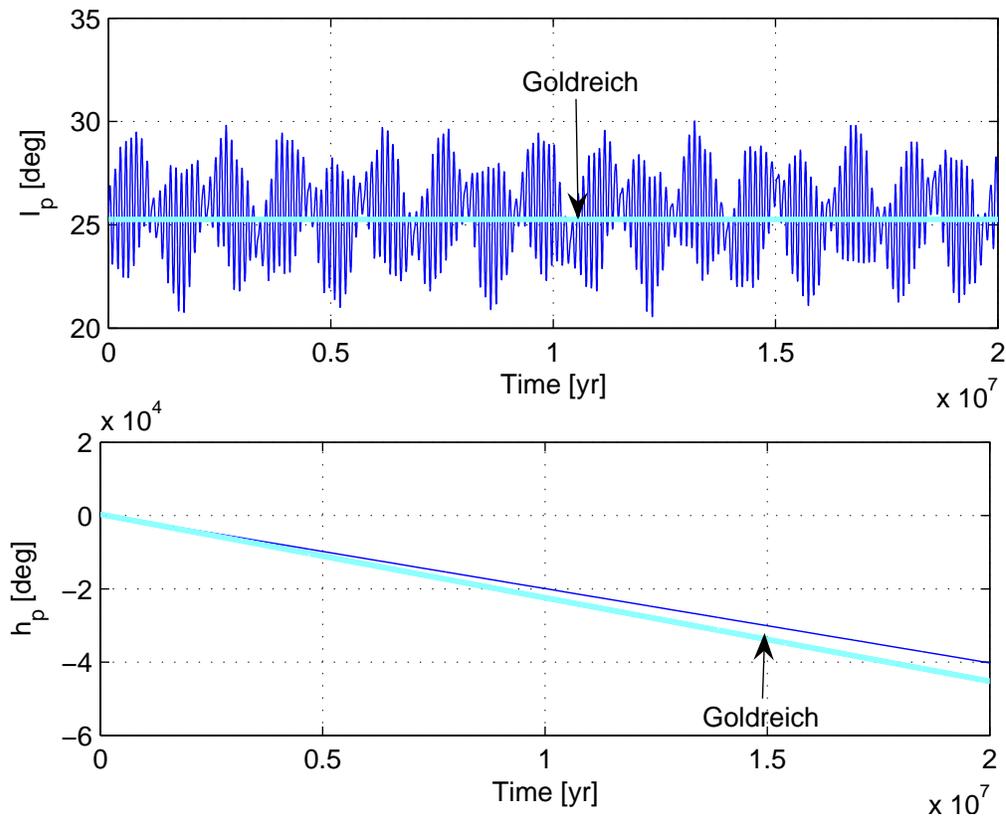}\\
  \caption{Preparation to comparison of the semianalytical model to Goldreich's model.
  In Goldreich's model, the Martian equator is assumed to precess uniformly, while the
  semianalytical model takes into account (via Colombo's equation) variations of the
  equinoctial precession.}
  \label{fig:goldreich1}
\end{figure}
\end{center}
\begin{center}
\begin{figure}[h]
  \includegraphics[width=6.0in]{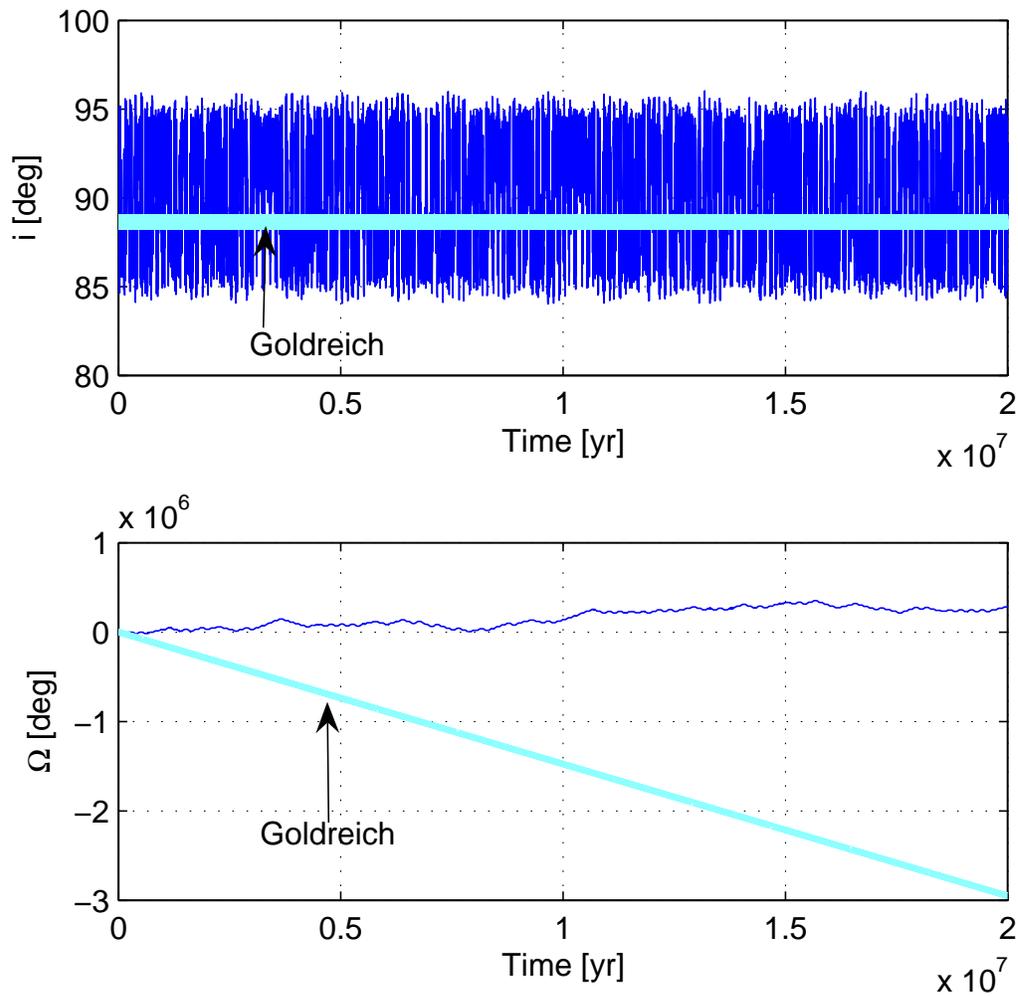}\\
  \caption{Comparison of the semi-analytical model to Goldreich's model
  for a satellite of Deimos' mass and with the initial conditions given
  by (\ref{42}).
  While in Goldreich's model the inclination remains tightly locked,  the
  semianalytical model reveals much larger variations of $\,\inc\,$.}
  \label{fig:goldreich2}
\end{figure}
\end{center}

 The final step in our study will be to compare the semianalytical model to
 Goldreich's approximation (\ref{25} - \ref{29}). To that end, we integrate
 our semianalytical model for 20 Myr, using the initial conditions from
 Table 2 with $\,i_0\,=\,89\,$ deg; and compare the outcome with that
 resulting from Goldreich's approximation simulated with the same initial
 conditions. The results of this comparison are depicted in
 Figs.~\ref{fig:goldreich1} - \ref{fig:goldreich2}. Specifically,
 Fig.~\ref{fig:goldreich1} compares the time histories of $\,I_p\,$ and
 $\,h_p\,$. There are noticeable differences in the dynamics of $\,I_p\,$.
 While Goldreich's approximation assumes a constant $\,I_p\,$, the
 semianalytical model is based on the Colombo calculation of the equinoctial
 precession, calculation that predicts considerable oscillations within
 the range $\,21\,\deg\,\le\,I_p\,\le\,30\,\deg\,$. Beside this, in our
 semianalytical model we take into account the direct gravitational pull
 exerted by the Sun on the satellite. All this entails differences between
 the dynamics predicted by our semianalytical model and the dynamics stemming
 from the Goldreich approximation. These differences, for $\,\inc\,$ and
 $\,\Omega\,$, are depicted in Fig.~\ref{fig:goldreich2}. In Goldreich's
 model, $\,\inc\,$ stays very close to the initial value:
 $\,88.27\,\deg\,\le\,\inc\,\le\,89.01\,\deg\;$, a behaviour that makes the
 essence of the \emph{Goldreich lock}. However, in the more accurate,
 semianalytical model we have $\,84\,\deg\,\le\,\inc\,\le\,96\,\deg\,$. The
 time history of $\,\Omega\,$, too, reveals that Goldreich's approximation
 does not adequately model the actual dynamics, since it predicts a much
 larger secular change than the semi-analytical model.

 All in all, the dynamics (i.e., particular trajectories) predicted by the two
 models are quantitatively different.
 At the same time, when it comes to the most physically important conclusion
 from the Goldreich approximation, the ``Goldreich lock" of the inclination,
 one may still say that, qualitatively, the semianalytical model confirms
 the locking even in the case when the equinoctial precession is variable,
 and the solar pull on the satellite is included. The locking survives even
 for highly inclined satellite orbits. It is, though, not as stiff as
 predicted by the Goldreich model: we can see from Fig.~\ref{fig:goldreich2}
 that the orbit inclination varies within a five-degree span, while the
 Goldreich approximation would constrain it to fractions of a degree.

 An intriguing fine feature of the inclination evolution, which manifests
 itself for orbits close to polar, is the ``crankshaft". This kind of
 behaviour, well defined in Fig.~\ref{fig:goldreich2}, is not rendered by the
 Goldreich model, because that model was initially developed for small
 inclinations and uniform precession. One might suspect that the
 ``crankshaft" is merely a numerical
 artefact, had it not been discovered under different circumstances (in the
 absence of precession but in the presence of a third body) by Zhang \&
 Hamilton (2005). This kind of pattern may be generic for the close
 vicinity of $\,\inc\,=\,90\,\deg\,$.


 \section{Conclusions}

 In the article thus far, we continued developing a tool for exploring
 long-term evolution of a satellite orbit about a precessing oblate primary.
 In particular, we were interested in the time-dependence of the orbit
 inclination relative to the moving equator of date. Our model includes
 three factors: $\,J_2\,$ of the planet, the planet's nonuniform equinoctial
 precession described by the Colombo formalism, and the gravitational pull
 exerted by the Sun on the satellite. The problem was approached using
 different methods. One, semianalytical, was based on numerical integration
 of the averaged Lagrange-type equations for the secular parts of the
 Keplerian orbital elements introduced in a noninertial reference frame
 coprecessing with the planetary equator of date. The right-hand sides of
 these equations consisted of precession-generated contributions and
 contributions due to the direct pull of the Sun. The other approach was
 a straightforward, purely numerical, computation of the satellite dynamics
 using Cartesian coordinates in a quasi-inertial reference frame.

 The results of both computations have demonstrated a good agreement over
 10 Myr. This means that the semianalytical model adequately describes the
 dynamics over this time span. Specifically, the terms neglected in the
 semianalytical model (the short-period terms and the terms of order
 $\,O(\mubold^2)\,$) play no significant role on this time scale.

 Our calculations have shown the advantages and the limitations of a simple
 model developed by Goldreich (1965) for uniform equinoctial precession and
 low inclinations. Though his model does not adequately describe the dynamics
 (that turns out to be chaotic), the main physical prediction of Goldreich's
 model -- the ``Goldreich lock" -- sustains the presence of the Sun and
 variations of equinoctial precession, provided the initial inclination is
 sufficiently low. For low initial inclinations, the inclination exhibits
 variations of order fractions of a degree. For higher inclinations, however,
 it varies already within a span of about ten degrees.
 For near-polar orbits, the inclination behaviour demonstrates the
 ``crankshaft", an chaotic pattern not accounted for by the Goldreich model.
 The ``crankshaft" emerges because in our model both the precession variations
 and the pull of the Sun are included into the model. However, numerical
 experiments have also shown that the ``crankshaft" gets generated by each of
 these two factors separately.



\section*{Appendix A: Closed-Form Expressions for $\dot \mu_1,\, \dot \mu_2 ,\, \dot\mu_3$}

Using the compact notation
$\,\cc_{(\cdot)}\,\equiv\,\cos(\cdot)\,$ and
$\,\s_{(\cdot)}\,\equiv\,\sin(\cdot)\,$, and conforming to the
procedure described in the text, we obtain the following
expressions for $\;\dot \mu_1,\, \dot \mu_2 ,\, \dot\mu_3\;$:
\begin{eqnarray}
  \dot\mu_1 &=&  (  (  (  ( 6\,{p}^{2}{q}^{2}-{p}^{4}-{q}^{4}
 )  ( \cc_{I_p}  ) ^{3}+ ( -6
\,{p}^{2}{q}^{2}+{p}^{4}+{q}^{4} ) \cc_{I_p}
 )  ( \cc_{h_p}  ) ^{4} \nonumber \\ &+& (
 ( -3\,{p}^{4}-6\,{p}^{2}{q}^{2}-3\,{q}^{2}+3\,{p}^{2}+5\,{q}^{4}
 )  ( \cc_{I_p}  ) ^{3}+ ( 2
\,{p}^{4}-4\,{q}^{4}+6\,{p}^{2}{q}^{2}-2\,{p}^{2}+2\,{q}^{2} )
\cc_{I_p}  ) \nonumber \\ &\cdot& ( \cc ( { h_p}
 )  ) ^{2}+ ( 3\,{q}^{2}-3\,{p}^{2}{q}^{2}-4\,{q}^{4}
 )  ( \cc_{I_p}  ) ^{3}+ ( 3
\,{q}^{4}-2\,{q}^{2}+2\,{p}^{2}{q}^{2} ) \cc_{I_p}\nonumber \\&+& (
( ( 4\,q{p}^{3}-4\,{q}^{3}p )
 ( \cc_{I_p}  ) ^{3}+ ( 4\,{q}^{3}p
-4\,q{p}^{3} ) \cc_{I_p}  )  ( \cc_{h_p} ) ^{5} + ( ( 2\,q{p}^{3}
-6\,qp+14\,{q}^{3}p ) ( \cc_{I_p}
 ) ^{3} \nonumber \\ &+& ( -12\,{q}^{3}p+4\,qp ) \cc_{I_p}  )  ( \cc_{h_p}  ) ^{3}+
 (  ( 6\,qp-10\,{q}^{3}p-6\,q{p}^{3} )  ( \cc_{ I_p}   ) ^{3}+ ( 4\,q{p}^{3}+8\,{q}^{3}p
-4\,qp ) \cc_{I_p}  ) \cc_{h_p} )\nonumber \\ &\cdot&(\s_{ h_p}
)^{-1}  ) ( \s_{ I_p}   ) ^{-1}+ (  (  ( -9\,p{
q}^{2}+9\,p{q}^{4}-3\,{p}^{5}+3\,{p}^{3}+6\,{p}^{3}{q}^{2} )
 ( \cc_{I_p}  ) ^{2}+3\,p{q}^{2}-2\,{p}^
{3}{q}^{2}+{p}^{5}\nonumber \\ &-&3\,p{q}^{4}-{p}^{3} )  ( \cc_{
h_p}  ^{3}+ (  ( -11\,{p}^{3}{q}^{2}-{p}^
{5}-10\,p{q}^{4}-p+2\,{p}^{3}+11\,p{q}^{2} )  ( \cc
 ( { I_p} )  ) ^{2}+3\,p{q}^{4}+3\,{p}^{3}{q}^{2}\nonumber \\ &-&3
\,p{q}^{2} ) \cc_{h_p} +( (
 ( -9\,{p}^{2}q+6\,{p}^{2}{q}^{3}+9\,{p}^{4}q-3\,{q}^{5}+3\,{q}^{
3} )  ( \cc_{I_p}  ) ^{2}-3\,{p}^{
4}q+3\,{p}^{2}q-2\,{p}^{2}{q}^{3}\nonumber \\ &-&{q}^{3}+{q}^{5} ) (
\cc_{ h_p}  ^{4}+ (  ( 7\,{p}^{2}q-{p}
^{2}{q}^{3}-8\,{q}^{3}-8\,{p}^{4}q+7\,{q}^{5}+q )  ( \cc_{ I_p}
) ^{2}-2\,{q}^{5}+{p}^{2}{q}^{3}+2\,{q }^{3}\nonumber \\
&+&3\,{p}^{4}q-3\,{p}^{2}q )  ( \cc ( { h_p}
 )  ) ^{2}+ ( 5\,{q}^{3}-4\,{q}^{5}-q-5\,{p}^{2}{q}^{
3}-{p}^{4}q+2\,{p}^{2}q )  ( \cc_{I_p}
 ) ^{2}-{q}^{3}+{q}^{5}+{p}^{2}{q}^{3})\nonumber \\ &\cdot&(\s ( { h_p}
 ) )^{-1}  ) \left[{ \sqrt{1-p^2-q^2}}\right]^{-1} ) {\alpha}^{2} - \alpha (
 (  (  ( -2\,q{ \dot{p}}-2\,{ \dot{q}}\,p )
 ( \cc_{I_p}  ) ^{2}+2\,q{ \dot{p}}+2
\,{ \dot{q}}\,p )  ( \cc_{h_p}
 ) ^{2}\nonumber \\ &+& ( { \dot{q}}\,p+q{ \dot{p}} )  (
\cc_{I_p}  ) ^{2}-{ \dot{q}}\,p-q{ \dot{p}}+( (  ( -2\,p{
\dot{p}}+2\,q{ \dot{q}}
 )  ( \cc_{I_p}  ) ^{2}+2\,p{
\dot{p}}-2\,q{ \dot{q}} )  ( \cc_{h_p}
 ) ^{3}+ (  ( 2\,p{ \dot{p}}-2\,q{ \dot{q}}
 )  ( \cc_{I_p}  ) ^{2}\nonumber \\ &-&2\,p{
\dot{p}}+2\,q{ \dot{q}} ) \cc_{h_p} )(\s
 ( { h_p} ) )^{-1} )  ( \s ( { I_p}
 )  ) ^{-1}+ (  ( -{ \dot{q}}\,{p}^{2}-2\,{q}^
{2}{ \dot{q}}+{ \dot{q}}-qp{ \dot{p}} ) \cc_{I_p} \cc_{h_p}
\nonumber \\ &+& {\frac { ( { \dot{p}} -2\,{p}^{2}{ \dot{p}}-pq{
\dot{q}}-{ \dot{p}}\,{q}^{2} ) \cc_{I_p}  ( \cc_{h_p}
 ) ^{2}+ ( -{ \dot{p}}+{ \dot{p}}\,{q}^{2}+2\,{p}^{2}{
 \dot{p}}+pq{ \dot{q}} ) \cc_{I_p}
}{\s_{ h_p}  }} ) \left[{ \sqrt{1-p^2-q^2}}\right]^{-1} )
~~~,~~~~~~~~~~~~
\end{eqnarray}
\begin{eqnarray}
  \dot\mu_2 &=& (  (  (  ( 4\,{q}^{3}p-4\,q{p}^{3} ) (\cc_{I_p}) ^{2}+4\,q{p}^{3}-4\,{q}^
{3}p )  ( \cc_{h_p}  ) ^{4}
  \nonumber \\
   &+&
 (  ( q{p}^{3}-7\,{q}^{3}p+2\,qp )  ( \cc_{ I_p}   ) ^{2}-3\,q{p}^{3}+5\,{q}^{3}p
 )  ( \cc_{h_p}  ) ^{2}+ ( 2
\,{q}^{3}p-qp+q{p}^{3} )  ( \cc_{I_p}
 ) ^{2} \nonumber \\ &-& {q}^{3}p+ ( (  ( 6\,{p}^{2}{q}^{2}-{p}^{
4}-{q}^{4} )  ( \cc_{I_p}  ) ^{2}- 6\,{p}^{2}{q}^{2}+{p}^{4}+{q}^{4}
)  ( \cc_{ h_p}  ^{5}+ (  ( 3\,{q}^{4}-{q}^{2}-9\,{p}^{2}{q
}^{2}+{p}^{2} )  ( \cc_{I_p}  ) ^{ 2} \nonumber
\\ &-& {p}^{4}-2\,{q}^{4}+9\,{p}^{2}{q}^{2} )  ( \cc_{h_p} ) ^{3}+ ( (
-{p}^{2}+{q}^{2}-2\,{q}^ {4}+{p}^{4}+3\,{p}^{2}{q}^{2} ) ( \cc (
{I_p} ))^{2}+{q}^{4} \nonumber \\ &-& 3\,{p}^{2}{q}^{2} ) \cc ( {
h_p} ) ) (\s_{ h_p}  ) ^{-1}   )  ( \s
 ( { I_p} )  ) ^{-1}+ (  ( -4\,{p}^{2}{q}
^{3}+6\,{p}^{2}q-2\,{q}^{3}+2\,{q}^{5}-6\,{p}^{4}q ) \cc
 ( { I_p} )  ( \cc_{h_p}
 ) ^{3} \nonumber \\ &+& ( 2\,{q}^{3}+4\,{p}^{4}q+2\,{p}^{2}{q}^{3}-4\,{p}^
{2}q-2\,{q}^{5} ) \cc_{I_p} \cc_{h_p} + ( (
-2\,{p}^{5}-6\,p{q}^{2}+6\,p{q}^{4}+4\,{p }^{3}{q}^{2}+2\,{p}^{3} )
\cc_{I_p}  ( \cc_{h_p}  ) ^{4} \nonumber\\ &+& ( 2\,{p}^{5}+8\,p{q}^
{2}-8\,p{q}^{4}-6\,{p}^{3}{q}^{2}-2\,{p}^{3} ) \cc_{I_p} ( \cc_{h_p}
) ^{2}+
 ( 2\,p{q}^{4}-2\,p{q}^{2}+2\,{p}^{3}{q}^{2} ) \cc_{I_p} )(\s_{ h_p}  )^{-1}  ) \nonumber \\
 &\cdot& \left[{ \sqrt{1-p^2-q^2}}\right]^ {-1} )
{\alpha}^{2} - \alpha (  (  ( ( 2\,p{ \dot{p}}-2\,q{ \dot{q}} )  (
\cc_{I_p}
 ) ^{3}+ ( -2\,p{ \dot{p}}+2\,q{ \dot{q}} ) \cc_{ I_p}   )  ( \cc_{ h_p}  ^{2} \nonumber \\
 &+& ( 2\,p{ \dot{p}}+4\,q{ \dot{q}}
 )  ( \cc_{I_p}  ) ^{3}+ ( -2
\,p{ \dot{p}}-4\,q{ \dot{q}} ) \cc_{I_p} + ( (  ( -2\,q{
\dot{p}}-2\,{ \dot{q}}\,p )
 ( \cc_{I_p}  ) ^{3} +( 2\,q{
\dot{p}}+ 2\,{ \dot{q}}\,p ) \cc_{I_p}
 )  ( \cc_{h_p}  ) ^{3} \nonumber \\ &+&  (
 ( 2\,q{ \dot{p}}+2\,{ \dot{q}}\,p )  ( \cc
 ( { I_p} )  ) ^{3}+ ( -2\,q{ \dot{p}}-2\,{
 \dot{q}}\,p ) \cc_{I_p}  ) \cc
 ( { h_p} ) )(\s_{ h_p}  )^{-1}  )
 ( \s_{ I_p}   ) ^{-1} \nonumber \\ &+&  (  (
 ( 2\,{ \dot{p}}\,{q}^{2}+2\,pq{ \dot{q}}+4\,{p}^{2}{
\dot{p}}-2\,{ \dot{p}} )  ( \cc_{I_p}
 ) ^{2}+{ \dot{p}}-2\,{p}^{2}{ \dot{p}}-pq{ \dot{q}}-{
\dot{p}}\,{q}^{2} ) \cc_{h_p}  \nonumber \\ &+&   (
 (  ( -2\,qp{ \dot{p}}-2\,{ \dot{q}}\,{p}^{2}-4\,{q}^{2}
{ \dot{q}}+2\,{ \dot{q}} )  ( \cc_{ I_p}   ) ^{2}-{ \dot{q}}+{
\dot{q}}\,{p}^{2}+2\,{q}^{2}{
 \dot{q}}+qp{ \dot{p}})(\cc_{ h_p}^{2}\nonumber\\&+&(2\,qp{\dot{p}}-2\,{\dot{q}}+2\,{
 \dot{q}}\,{p}^{2}+4\,{q}^{2}{ \dot{q}})(\cc_{I_p})^{2}-{\dot{q}}\,{p}^{2}-2\,{q}^{2
}{\dot{q}}+{ \dot{q}}-qp{ \dot{p}}) (\s ( { h_p}
 ) )^{-1}  ) \left[{ \sqrt{1-p^2-q^2}}\right]^{-1} )
 \nonumber ~~~,~~~~~~\\
\end{eqnarray}
and

\pagebreak

\begin{eqnarray}
 \dot  \mu_3 &=& (  (  (  ( 4\,{q}^{3}p-4\,q{p}^{3})(\cc_{I_p}) ^{2}+4\,q{p}^{3}-4\,
 {q}^{3}p )  ( \cc_{h_p}  ) ^{4}+
 (  ( q{p}^{3}-7\,{q}^{3}p+2\,qp )  ( \cc
 ( { I_p} )  ) ^{2} \nonumber \\ &-& 3\,q{p}^{3}+5\,{q}^{3}p
 )  ( \cc_{h_p}  ) ^{2}+ ( 2
\,{q}^{3}p-qp+q{p}^{3} )  ( \cc_{I_p}
 ) ^{2}-{q}^{3}p+ ( (  ( 6\,{p}^{2}{q}^{2}-{p}^{
4}-{q}^{4} )  ( \cc_{I_p}  ) ^{2}- 6\,{p}^{2}{q}^{2}+{p}^{4}
\nonumber \\ &+&{q}^{4} )  ( \cc ( { h_p}
 )  ) ^{5}+ (  ( 3\,{q}^{4}-{q}^{2}-9\,{p}^{2}{q
}^{2}+{p}^{2} )  ( \cc_{I_p}  ) ^{
2}-{p}^{4}-2\,{q}^{4}+9\,{p}^{2}{q}^{2} )  ( \cc_{h_p} ) ^{3}+ (  (
-{p}^{2}+{q}^{2}-2\,{q}^ {4}\nonumber \\
&+&{p}^{4}+3\,{p}^{2}{q}^{2} ) ( \cc ( { I_p}
 )  ) ^{2}+{q}^{4}-3\,{p}^{2}{q}^{2} ) \cc ( {
 h_p} ) )(\s_{ h_p}  )^{-1}  )  (
\s
 ( { I_p} )  ) ^{-1}+ (  ( -4\,{p}^{2}{q}
^{3}+6\,{p}^{2}q-2\,{q}^{3}+2\,{q}^{5}\nonumber \\ &-&6\,{p}^{4}q )
\cc
 ( { I_p} )  ( \cc_{h_p}
 ) ^{3}+ ( 2\,{q}^{3}+4\,{p}^{4}q+2\,{p}^{2}{q}^{3}-4\,{p}^
{2}q-2\,{q}^{5} ) \cc_{I_p} \cc_{h_p} \nonumber \\ &+& ( (
-2\,{p}^{5}-6\,p{q}^{2}+6\,p{q}^{4}+4\,{p }^{3}{q}^{2}+2\,{p}^{3} )
\cc_{I_p}  ( \cc_{h_p}  ) ^{4}+ ( 2\,{p}^{5}+8\,p{q}^
{2}-8\,p{q}^{4}-6\,{p}^{3}{q}^{2}-2\,{p}^{3} ) \cc_{I_p} ( \cc_{h_p}
) ^{2}\nonumber \\ &+&
 (2\,p{q}^{4}-2\,p{q}^{2}+2\,{p}^{3}{q}^{2})\cc_{I_p})(\s_{ h_p})^{-1}){{
 \sqrt{1-p^2-q^2}}}^ {-1} )
{\alpha}^{2} - \alpha (  (  ( ( 2\,p{ \dot{p}}-2\,q{ \dot{q}} )  (
\cc_{I_p}
 ) ^{3}\nonumber \\ &+& ( -2\,p{ \dot{p}}+2\,q{ \dot{q}} ) \cc
 ( { I_p} )  )  ( \cc ( { h_p}
 )  ) ^{2}+ ( 2\,p{ \dot{p}}+4\,q{ \dot{q}}
 )  ( \cc_{I_p}  ) ^{3}+ ( -2
\,p{ \dot{p}}-4\,q{ \dot{q}} ) \cc_{I_p} \nonumber \\ &+& ( (  (
-2\,q{ \dot{p}}-2\,{ \dot{q}}\,p )
 ( \cc_{I_p}  ) ^{3}+ ( 2\,q{
\dot{p}}+2\,{ \dot{q}}\,p ) \cc_{I_p}
 )  ( \cc_{h_p}  ) ^{3}+ (
 ( 2\,q{ \dot{p}}+2\,{ \dot{q}}\,p )  ( \cc
 ( { I_p} )  ) ^{3}\nonumber \\ &+& ( -2\,q{ \dot{p}}-2\,{
 \dot{q}}\,p ) \cc_{I_p}  ) \cc
 ( { h_p} ) )(\s_{ h_p}  )^{-1}  )
 ( \s_{ I_p}   ) ^{-1}+ (  (
 ( 2\,{ \dot{p}}\,{q}^{2}+2\,pq{ \dot{q}}+4\,{p}^{2}{
\dot{p}}-2\,{ \dot{p}} )  ( \cc_{I_p}
 ) ^{2}+{ \dot{p}}\nonumber \\ &-&2\,{p}^{2}{ \dot{p}}-pq{ \dot{q}}-{
\dot{p}}\,{q}^{2} ) \cc_{h_p} + (
 (  ( -2\,qp{ \dot{p}}-2\,{ \dot{q}}\,{p}^{2}-4\,{q}^{2}
{ \dot{q}}+2\,{ \dot{q}} )  ( \cc ( { I_p}
 )  ) ^{2}-{ \dot{q}}+{ \dot{q}}\,{p}^{2}+2\,{q}^{2}{
 \dot{q}}\nonumber \\ &+&qp{ \dot{p}})(\cc_{ h_p})^{2}+( 2\,qp{\dot{p}}-2\,{\dot{q}}+2\,{
 \dot{q}}\,{p}^{2}+4\,{q}^{2}{\dot{q}})(\cc_{I_p})^{2}-{\dot{q}}\,{p}^{2}-2\,{q}^{2
}{ \dot{q}}+{ \dot{q}}-qp{ \dot{p}})(\s_{ h_p}  )^{-1}  ) \nonumber
\\ &\cdot&\left[{ \sqrt{1-p^2-q^2}}\right]^{-1} )
\end{eqnarray}

\section*{Appendix B: Niching Genetic Algorithms}

 The most commonly used Genetic Algorithm (GA) is the so-called ``Simple GA." To
 perform an evolutionary search, the Simple GA uses the operators of crossover,
 reproduction, and mutation. A crossover is used to create new solution strings
 (``children" or ``offspring") from the existing strings (``parents").
 Reproduction copies individual strings according to the objective function
 values. Mutation is an occasional random alteration of the value of a string
 position, used to promote diversity of solutions.

 Although Simple GA's are capable of detecting the global optimum, they suffer from
 two main drawbacks. First, convergence to a local optimum is possible due to the
 effect of premature convergence, where all individuals in a population become
 nearly identical before the optima has been located. Second, convergence to a
 single optimum does not reveal other optima, which may exhibit attractive features.
 To overcome these problems, modifications of Simple GA's were considered. These
 modifications are called niching methods, and are aimed at promoting a diversity
 of solutions for multi-modal optimisation problems. In other words, instead of
 converging to a single (possibly local) optimum, niching allows for a number of
 optimal solutions to co-exist, and it lets the designer choose the appropriate one.
 The niching method used throughout this study is that of Deterministic Crowding.
 According to this method, individuals are first randomly grouped into parent pairs.
 Each pair generates two children by application of the standard genetic operators.
 Every child then competes against one of his parents. The winner of the competition
 moves on to the next generation. By using the notation $\;P_{i}\;$ for a parent, $
 \;C_{i}\;$ for a child, $\;f(\cdot )\;$ for a fitness, and $\;d(\cdot)\;$ for a
 distance, a pseudo-code for the two possible parent-child tournaments can be written
 as follows:\\

 If  $\left[d(P_{1} ,\, \, C_{1} )+d(P_{2} ,\, \, C_{2}
 )=d(P_{1} ,\, \, C_{2} )+d(P_{2} ,\, \, C_{1} )\right]$

 If  $f(C_{1} )\ge f(P_{1} )$  replace  $P_{1} $  with  $C_{1} $

 If  $f(C_{2} )\ge f(P_{2} )$  replace  $P_{2} $  with  $C_{2} $

 Else\\

 If  $f(C_{1} )\ge f(P_{2} )$  replace  $P_{2} $  with  $C_{1} $

 If  $f(C_{2} )\ge f(P_{1} )$  replace  $P_{1} $  with  $C_{2} $\\

 In addition to applying the Deterministic Crowding niching method, we used a two-point
 crossover instead of a single-point one. In the Simple GA, the crossover operator breaks the
 binary string of parameters, the ``chromosome," at a random point and exchanges the two
 pieces to create a new ``chromosome." In a two-point crossover, the ``chromosome" is
 represented with a ring. The string between the two-crossover points is then exchanged.
 The two-point crossover or other multiple-point crossover schemes have preferable
 properties when optimisation highly nonlinear functions is performed.

\section*{Acknowledgments}

 ME is grateful to George Kaplan for numerous fruitful discussions on the subject, and to
 Marc Murison for a consultation on the possible scenario of the Martian satellites capture.
 The work of ME was partially supported by NASA grant W-19948. The work of VL was supported by
 the European Community's Improving Human Potential Programme contract RTN2-2001-00414
 MAGE.

{}

 ~\\


 \section*{Figures}

 ~\\

 ~\\

%

%

%
%
%
%

%
%
%
%

\end{document}